\documentclass[article]{aa} % for a referee version
\usepackage{graphicx}

\def\x2{$\chi^{2}$}

\def\asca{{\it ASCA}}

\def\xmm{{\it XMM-Newton}}
\def\chandra{{\it Chandra }}

\def\lunits{$\rm erg~s^{-1}~$}
\def\funits{$\rm erg~cm^{-2}~s^{-1}~$}
\def\cunits{$\rm cm^{-2}~$}

\begin{document}

   \title{{\xmm} and \chandra measurements of the AGN intrinsic absorption:
   dependence on luminosity and redshift}

   \titlerunning{The AGN intrinsic absorption}
    \authorrunning{A. Akylas}

   \author{A. Akylas
          \inst{1},
          I. Georgantopoulos
	  \inst{1},
	  A. Georgakakis 
	  \inst{2},
	  S. Kitsionas
	   \inst{3},
  	  and
	  E. Hatziminaoglou
	   \inst{4}
          }

   \offprints{A. Akylas}

   \institute{$^1$Institute of Astronomy \& Astrophysics, National Observatory 
 of Athens, I.Metaxa \& B. Pavlou, Penteli, 15236, Athens, Greece \\
 $^2$Physics Department, Imperial College of Science Technology and Medicine, 
Blackett Laboratory, Prince Consort Rd, 2BZ SW7, London, U.K. \\
  $^3$Astrophysikalisches Institut Potsdam, An der Sternwarte 16, D-14482, Potsdam, Germany  \\
  $^4$Institute de Astrofisica de Canarias, C/Via Lactea s/n, E-38200 La Laguna, Spain  \\
     }   
       \date{}

   \abstract
   {We combine bright {\xmm} data  with the 
   \chandra Deep Field South observations in order to explore the 
   behavior of the intrinsic AGN absorption,  
   as a function  of redshift and luminosity.  
   Our sample consists of 359 sources selected in the hard 2-8 keV 
   band, spanning the flux range $6\times 10^{-16}$-$3\times10^{-13}$ 
   \funits with a high rate of spectroscopic or photometric redshift 
   completeness (100 and 85 per cent respectively for the 
   {\chandra} and {\xmm} data). We derive the column density values using
   X-ray spectral fits. We find that the fraction of obscured AGN falls  
   with increasing luminosity in agreement with previous findings.  
   The fraction of obscured AGN shows an apparent increase at 
   high redshifts ($z>2$). Simulations show that this effect can be 
   most probably attributed to the fact that at high redshifts the 
   column densities are overestimated.   
   {\keywords{Surveys -- X-rays: galaxies -- X-rays: general}}}

   \maketitle

\section{INTRODUCTION}
 
Ultra-deep surveys with {\chandra} have resolved the bulk of the X-ray
background at hard energies (2-10\,keV), shedding light on the nature
of the AGN population at faint fluxes  ($f_X(2-8~\rm keV) \sim
1.4\times10^{-16}$ \funits; Mushotzky et al. 2000, Giacconi et
al. 2002, Alexander et al. 2003). These surveys have revealed luminous
unobscured QSOs but also more importantly, large numbers of obscured  
AGN ($N_H>10^{22}$ \cunits). The fraction of absorbed sources rises
steeply with decreasing flux (e.g. Alexander et al. 2003) dominating
the X-ray population at the flux limit of the 2\,Ms \chandra Deep Field
North (CDF-N) survey (e.g. Perola et al. 2004). To the first approximation,  
this trend is in qualitative agreement with the predictions of the
X-ray background (XRB) population synthesis models (e.g. Comastri et
al. 1995; Gilli et al. 2001). A more detailed quantitative comparison
however, reveals a number of inconsistencies. For example, the models
above also predict large numbers of luminous ($L_X > 10^{44} \rm erg
\, s^{-1}$) heavily obscured ($N_H>10^{22} \, \rm cm^{-2}$) type-II
QSOs, although only a handful of such objects have been identified
todate, despite painstaking efforts (e.g. Barger et al. 2003; Fiore et
al. 2003; Szokoly et al. 2004). Parallel to the 
\chandra deep fields, surveys with the {\xmm}, probing on average
brighter fluxes, also provide a wealth of complementary information on
the nature and the  evolution of the AGN population. These brighter
surveys also show a clear scarcity of obscured AGN relative to the
model expectations (e.g. Piconcelli et al. 2002, Georgantopoulos et  
al. 2004, Perola et al. 2004). 

Although the source of the inconsistency between observations and model 
predictions remains unclear, two main scenarios are put forward to
reconcile the problem. The first  argues that observational biases are
affecting our conclusions, while the second proposes major revision of
the basic assumptions of the XRB models.  In the former case, it
is proposed that obscured AGN are present in current surveys but lie
in poorly explored regions of the parameter space. For example a large
fraction of X-ray sources, particularly in the \chandra deep
fields (about 30\%; Alexander et al. 2003) are optically faint ($R >
24.5$\,mag) and therefore, hard to study in detail. Treister et
al. (2004) argue that it is precisely this poorly explored population
that comprises a large fraction of obscured AGN at high redshift. This
is because although the energetic X-ray photons of these systems can
penetrate the large obscuring columns of gas and dust, the rest-frame
UV/optical light is heavily extincted resulting in faint observed
optical magnitudes. Consequently the type-II AGN are difficult
spectroscopic targets, even with 10-m class telescopes, and their
intrinsic properties (including their $N_H$) remain ill constrained.

In the case of the model revision scenario, one of the key input
parameters to the XRB models is the intrinsic column density ($N_H$)
distribution of AGN. This is assumed to be that of Seyfert galaxies
(i.e. low-luminosity AGN) measured from the local Universe (Risaliti
et al. 1999). It is possible that the luminous AGN identified in
deeper surveys do not follow the same column density distribution,
leading to the failure of the population synthesis models. Indeed,
Ueda et al. (2003) by combining sources with redshift information from
the CDF-N and  brighter {\asca} surveys find that the fraction of
obscured AGN diminishes with increasing luminosity. The physical
interpretation of this model could be that the highly luminous AGN
blow away the obscuring screen or they photoionize the gas  around
them.  More recently, La Franca et al. (2005) estimated the fraction
of absorbed AGN as a function of luminosity combining data from the
CDF-N and CDF-S as well as  bright {\xmm} data. They also suggest a
decrease of the obscured AGN fraction (hereafter F) with luminosity as
well as a possible increase with redshift, in contrast to the 
standard model assumption that F is independent of $L_X$ or $z$.  
Such studies are
obviously important for furthering our understanding of the AGN
unification models and  moreover, they are crucial for the
construction of X-ray background population synthesis models.    

 In this paper we attempt to further 
 investigate the intrinsic absorption in AGN,
 using a large X-ray sample from {\xmm} and {\chandra} 
 all with X-ray spectroscopic information.  
 In particular, we combine the  {\chandra} Deep Field South, CDF-S 
observations (162 sources, {\it all}  with  
redshift information, from  Szokoly et al. 2004 and Zheng et al. 2004),   
with {\xmm} data at brighter fluxes ($>3\times10^{-14}$ \funits), 
from  the HELLAS2XMM survey of Perola et al. (2004) (44 sources)  
and our own {\xmm} survey which overlaps  with the Sloan Digital 
Sky Survey, SDSS (153 sources).  
 The advantages of our approach are the following:

 $\bullet$ We have derived X-ray spectra for all the X-ray sources 
 in our sample, trying to avoid the uncertainties that can be 
 introduced by a simple X-ray colour.  

 $\bullet$ We estimate the fraction of 
 obscured AGN using 
 the $1/V_{m}$ method. In this way we properly 
   take into account the 'accessible' 
 volume of each source, i.e. the fact that
 many obscured sources are not detected at 
 a given flux limit.  
   
 $\bullet$ We use the CDF-S, instead of the CDF-N, as the former
 has photometric or spectroscopic redshifts for all the sources. 
 This means that there are no missing type-II 
 AGN at high redshifts which would introduce a spurious 
 correlation between obscured fraction and luminosity. 

Throughout this paper we adopt H$_{\circ}$=70 km s$^{-1}$ Mpc$^{-1}$, 
$\Omega_m$=0.3 and $\Lambda$=0.7

\section{THE X-RAY DATA}
\subsection{The {\xmm} data}

The {\xmm} X-ray data come from 28 public fields 
selected to overlap with the second data release 
of the SDSS (DR2; Stoughton et al. 2002) and cover a 
total area of $\sim 5$ sq. degrees. Eight of 
these fields comprise the North {\xmm}/2dF 
survey (see for details Georgakakis 
et al. 2003, Georgantopoulos et al. 2004). For the 
fields observed more than once with {\xmm}, we use the 
deeper of the multiple observations. The observational 
characteristics of the 28 {\xmm} fields used here, 
obtained from Georgantopoulos \& Georgakakis (2005), are listed in 
Table \ref{table1}.

\begin{table*} 
\centering
\caption{The 28 archival {\xmm} pointings used in this study.}
\label{table1}
\begin{tabular}{ccccccc}

\hline 
RA & Dec & Filter & ${N_H}_{gal}$ & PN exp. time & MOS exp. time & Field Name \\
(J2000) & (J2000) & FILTER & ($\times 10^{20}$ cm$^{-2}$) & (sec) & (sec) &        \\
\hline
\hline
$23\mathrm{^h} 54\mathrm{^m} 09.0\mathrm{^s}$ & $  -10\mathrm{^\circ} 24\mathrm{^\prime} 00\mathrm{^{\prime\prime}}$ & MEDIUM & 2.91 & 13\,600 & 19\,100 & ABELL\,2670 \\
$23\mathrm{^h} 37\mathrm{^m} 40.0\mathrm{^s}$ & $  +00\mathrm{^\circ} 16\mathrm{^\prime} 33\mathrm{^{\prime\prime}}$ &  THIN   & 3.82 & 8\,200  & 13\,300 & RXCJ\,2337.6+0016 \\
$17\mathrm{^h} 01\mathrm{^m} 23.0\mathrm{^s}$ & $  +64\mathrm{^\circ} 14\mathrm{^\prime} 08\mathrm{^{\prime\prime}}$ & MEDIUM & 2.65 & 2\,300  & 3\,900  & RXJ\,1701.3 \\
$15\mathrm{^h} 43\mathrm{^m} 59.0\mathrm{^s}$ & $  +53\mathrm{^\circ} 59\mathrm{^\prime} 04\mathrm{^{\prime\prime}}$ & THIN   & 1.27 & 14\,200 & 19\,200 & SBS\,1542+541 \\
$13\mathrm{^h} 49\mathrm{^m} 15.0\mathrm{^s}$ & $  +60\mathrm{^\circ} 11\mathrm{^\prime} 26\mathrm{^{\prime\prime}}$ & THIN   & 1.80 & 14\,100 & 18\,100 & NGC\,5322 \\
$13\mathrm{^h} 04\mathrm{^m} 12.0\mathrm{^s}$ & $  +67\mathrm{^\circ} 30\mathrm{^\prime} 25\mathrm{^{\prime\prime}}$ & THIN   & 1.80 & 14\,600 & 17\,100 & ABELL\,1674 \\
$12\mathrm{^h} 45\mathrm{^m} 09.0\mathrm{^s}$ & $  +00\mathrm{^\circ} 27\mathrm{^\prime} 38\mathrm{^{\prime\prime}}$ & MEDIUM & 1.73 & 46\,300 & 55\,500 & NGC\,4666 \\
$12\mathrm{^h} 31\mathrm{^m} 32.0\mathrm{^s}$ & $  +64\mathrm{^\circ} 14\mathrm{^\prime} 21\mathrm{^{\prime\prime}}$ & THIN   & 1.98 & 26\,100 & 30\,100 & MS\,1229.2+6430 \\
$09\mathrm{^h} 35\mathrm{^m} 51.0\mathrm{^s}$ & $  +61\mathrm{^\circ} 21\mathrm{^\prime} 11\mathrm{^{\prime\prime}}$ & THIN   & 2.70 & 20\,400 & 33\,900 & UGC\,5051 \\
$09\mathrm{^h} 34\mathrm{^m} 02.0\mathrm{^s}$ & $  +55\mathrm{^\circ} 14\mathrm{^\prime} 20\mathrm{^{\prime\prime}}$ & THIN   & 1.98 & 23\,500 & 28\,500 & IZW\,18 \\
$09\mathrm{^h} 17\mathrm{^m} 53.0\mathrm{^s}$ & $  +51\mathrm{^\circ} 43\mathrm{^\prime} 38\mathrm{^{\prime\prime}}$ & MEDIUM & 1.44 & 15\,900 & 13\,600 & ABELL\,773 \\
$08\mathrm{^h} 31\mathrm{^m} 41.0\mathrm{^s}$ & $  +52\mathrm{^\circ} 45\mathrm{^\prime} 18\mathrm{^{\prime\prime}}$ & MEDIUM & 3.83 & 66\,800 & 73\,300 & APM\,08279+5255 \\
$03\mathrm{^h} 57\mathrm{^m} 22.0\mathrm{^s}$ & $  +01\mathrm{^\circ} 10\mathrm{^\prime} 56\mathrm{^{\prime\prime}}$ & THIN   &13.20 & 19\,100 & 21\,400 & HAWAII\,167 \\
$03\mathrm{^h} 38\mathrm{^m} 29.0\mathrm{^s}$ & $  +00\mathrm{^\circ} 21\mathrm{^\prime} 56\mathrm{^{\prime\prime}}$ & THIN   & 8.15 & 8\,900  & 6\,700  & SDSS\,033829.31+00215 \\
$03\mathrm{^h} 02\mathrm{^m} 39.0\mathrm{^s}$ & $  +00\mathrm{^\circ} 07\mathrm{^\prime} 40\mathrm{^{\prime\prime}}$ & THIN   & 7.16 & 38\,100 & 46\,900 & CFRS\,3H \\
$02\mathrm{^h} 56\mathrm{^m} 33.0\mathrm{^s}$ & $  +00\mathrm{^\circ} 06\mathrm{^\prime} 12\mathrm{^{\prime\prime}}$ & THIN   & 6.50 &  --     & 11\,600 & RX\,J0256.5+000\\		
$02\mathrm{^h} 41\mathrm{^m} 05.0\mathrm{^s}$ & $  -08\mathrm{^\circ} 15\mathrm{^\prime} 21\mathrm{^{\prime\prime}}$ & MEDIUM & 3.07 & 12\,300 & 15\,600 & NGC\,1052 \\
$01\mathrm{^h} 59\mathrm{^m} 50.0\mathrm{^s}$ & $  +00\mathrm{^\circ} 23\mathrm{^\prime} 41\mathrm{^{\prime\prime}}$ & MEDIUM & 2.65 & 3\,800  &  --     & MRK\,1014 \\
$01\mathrm{^h} 52\mathrm{^m} 42.0\mathrm{^s}$ & $  +01\mathrm{^\circ} 00\mathrm{^\prime} 43\mathrm{^{\prime\prime}}$ & MEDIUM & 2.80 & 5\,800  & 17\,200 & ABELL\,267 \\
$00\mathrm{^h} 43\mathrm{^m} 20.0\mathrm{^s}$ & $  -00\mathrm{^\circ} 51\mathrm{^\prime} 15\mathrm{^{\prime\prime}}$ & MEDIUM & 2.33 & 15\,700 &  --     & UM\,269 \\
$13\mathrm{^h} 41\mathrm{^m} 24.0\mathrm{^s}$ & $  +00\mathrm{^\circ} 24\mathrm{^\prime} 00\mathrm{^{\prime\prime}}$ & THIN & 2.00 &  5779  & 9974 & F864-1\\
$13\mathrm{^h} 43\mathrm{^m} 00.0\mathrm{^s}$ & $+00\mathrm{^\circ} 24\mathrm{^\prime} 00\mathrm{^{\prime\prime}}$ & THIN & 2.00 & 2958  & 6586 & F864-2 \\
$13\mathrm{^h} 44\mathrm{^m} 36.0\mathrm{^s}$ & $+00\mathrm{^\circ} 24\mathrm{^\prime} 00\mathrm{^{\prime\prime}}$ & THIN & 2.00 & 2187  & 7727 & F864-3\\
$13\mathrm{^h} 43\mathrm{^m} 00.0\mathrm{^s}$ & $+00\mathrm{^\circ} 00\mathrm{^\prime} 00\mathrm{^{\prime\prime}}$ & THIN & 2.00& 1693 & 4447 & F864-5 \\
$13\mathrm{^h} 44\mathrm{^m} 36.0\mathrm{^s}$ & $+00\mathrm{^\circ} 00\mathrm{^\prime} 00\mathrm{^{\prime\prime}}$ &THIN & 2.00 & 2766 & 6493 & F864-6 \\
$13\mathrm{^h}$ $41\mathrm{^m} 24.0\mathrm{^s}$ & $-00\mathrm{^\circ} 24\mathrm{^\prime} 00\mathrm{^{\prime\prime}}$ &THIN &2.00 & 3459 & 7139 & F864-7\\
$13\mathrm{^h} 43\mathrm{^m} 24.0\mathrm{^s}$ & $-00\mathrm{^\circ} 24\mathrm{^\prime} 00\mathrm{^{\prime\prime}}$ &THIN &2.00 &2109 & 7276 & F864-8 \\
$13\mathrm{^h} 44\mathrm{^m} 36.0\mathrm{^s}$ & $-00\mathrm{^\circ} 24\mathrm{^\prime} 00\mathrm{^{\prime\prime}}$ &THIN &2.00 & 4545 & 8330 & F864-9 \\	
\hline

\end{tabular} 	
\end{table*}     

The X-ray data have been analyzed using the Scientific 
Analysis Software (SAS v6.0). The pipeline event files, 
produced by the {\xmm} Science Center were screened for 
high particle background periods by rejecting time intervals 
with 0.5-10 keV count rates higher than 25 and 15 cts/s 
for the PN and the two MOS cameras respectively. The PN 
and MOS good time intervals are listed in Table \ref{table1}. 
The differences between the PN and the MOS exposure times 
are due to varying  start and end times of the individual 
observations. We consider the events corresponding to 
patterns 0-4 for PN and 0-12 for MOS instrument. 

In order to increase the signal-to-noise (S/N) ratio and to reach 
fainter fluxes, we have merged the MOS and the PN data into a 
single event file using the {\sl MERGE} task of SAS. The X-ray images 
have been extracted in the 2-8 keV (hard) energy band, for both 
the merged and the individual event files. 
We use the more sensitive (higher S/N ratio) merged image 
for source extraction and flux estimation while the individual PN 
and MOS images are used to extract the spectral files. The source 
detection has been performed using the {\sl EWAVELET} task of SAS with 
a detection threshold of 6{$\sigma$}. The choice of the threshold 
minimizes the number of spurious sources 
in the final catalogue without reducing the number of real detections. 
All the detected sources were carefully inspected, one by one to 
exclude spurious detections associated with CCD gaps, hot pixels or 
lying close to the edge of the detector's field of view. We also 
exclude from the final catalogue the target source of each {\xmm} 
and seven sources that appear extended on the {\xmm} EPIC images and are 
clearly associated with diffuse cluster emission. We estimate the 
observed flux of each source in a 18 arcsec aperture, adopting 
a power-law energy distribution model with $\Gamma$=1.8. 
The final X-ray catalogue comprises a total of 507 sources detected 
in the 2-8 keV merged (PN+MOS) images with a 6$\sigma$, 2-8 keV, 
flux limit of 5${\times}10^{-15}$ \funits 
(see Georgakakis et al. 2006a). 
For the purpose of this study we consider only the brighter 
hard X-ray sources with a 2-8 keV flux $f_X$(2-8)$>$3$\times 10^{-14}$ 
\funits (169 sources). This is to ensure sufficient 
photon-statistics to perform X-ray spectral analysis and to minimize 
the number of objects without optical identifications. 
The background subtracted photon count distribution of these sources 
is plotted in Fig. \ref{counts}. 
\subsection{The CDF-S data}

The CDF-S X-ray data are obtained from Giacconi et al. (2002). 
The data from these observations include 247 hard (2-10 keV) X-ray 
selected sources with complete spectroscopic or photometric redshift 
(Szokoly et al. 2004 and Zheng et al. 2004). For one source (\#261)
there is no redshift information and hence it is excluded from our 
analysis.

\begin{figure}
   \centering
   \includegraphics[width=8.5cm]{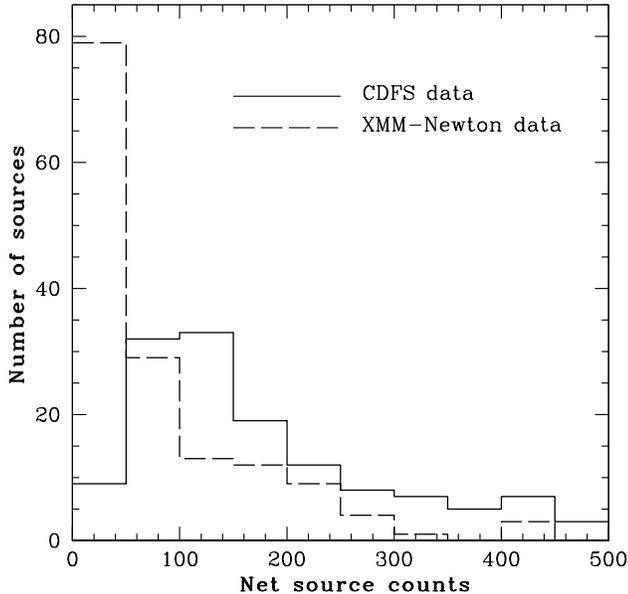}
   \caption{The background subtracted photon count 
	distribution for our {\xmm}/SDSS (dashed  
	histogram) and the CDF-S survey (solid histogram).}
    \label{counts}
    \end{figure}

These data expand the 2-8 keV flux coverage of our 
{\xmm} data down to $\sim$6$\times10^{-16}$ \funits. 
In order to increase the photon statistics we consider 
only the sources that lie within the central CDF-S region 
and therefore overlap with all 11 Chandra pointings. 
This reduces the number of the sources to 188. The vast 
majority of the sources (179) contain more than 50 net 
counts, sufficient for spectral analysis. Furthermore we 
exclude 26  sources identified as normal galaxy candidates 
by Norman et al. (2004). Therefore our final CDF-S AGN 
sample comprise 162 sources with spectroscopic (75 sources) 
or photometric (87 sources ) redshift information. The 
background subtracted photon count distribution of these 
sources is plotted in Fig. \ref{counts}.

\section{THE SAMPLE}
 
We use the SDSS DR2 catalogue to optically identify the hard 
X-ray detected sources using the method of Downes et al. (1986)
(for a detailed description of the followed procedure see
Georgakakis et al. 2004) to calculate the probability,$P$, a 
given candidate is a spurious identification. Here we apply
an upper limit in the search radius, $r<7$ arcsec and a cutoff
on the probability, $P<0.05$, to limit the optical identification
to those candidates that are least likely to be spurious alignments. 
The cross-correlation reveals 143 coincidences. 
26 sources have no optical counterpart in the SDSS limit ($r < 22.5$ mag) 
giving a 85 per cent completeness for our {\xmm} sample.
From the data above we further exclude all the sources with 
extended optical morphology and blue colors, i.e. $g-r<0.5$ mag, 
since the estimation of the photometric redshifts for these 
population is inaccurate as we discuss below. 
After this selection our catalogue comprises in total 157 
hard X-ray sources. In particular there are 45 sources with 
spectroscopic information, 86 with only photometric observations 
and 26 optically faint sources ($r>22.5$ mag).

For the 86 sources with photometric information available we 
estimate photometric redshifts using the method described by 
Kitsionas et al. (2005). These authors have applied the 
photometric redshift estimation technique of Hatziminaoglou,
Mathez \& Pell\'{o} (2000) on X-ray selected {\xmm} 
samples, also using 5-band photometry from the SDSS. The method 
is based on the standard $\chi^{2}$ minimization, using for the 
redshift estimation a combination of QSO stellar and galaxy 
Spectral Energy Distribution (SED) templates. In particular, 
the method uses the three QSO templates of Hatziminaoglou et 
al. (2000) produced by varying, between 0 and 1, the optical 
power-law spectral index of simulated QSO spectra that include 
a variety of broad emission lines. It also uses four different 
galaxy templates (E/S0, Sbc, Scd, Im) from  Coleman, Wu  \& 
Weedman (1980).

Following the Kitsionas et al. (2005) procedure we have divided the 
sources in our photometric sample into point-like and extended 
according to their SDSS optical morphology. We have estimated 
photometric redshifts using the QSO templates for the point-like 
objects and the galaxy templates for the extended sources 
respectively. As explained above, we have excluded extended objects 
with blue colours ($g-r<0.5$), as Kitsionas et al. (2005) have 
shown that the photometric redshift estimations for such objects 
are unreliable using template fitting techniques with the existing 
QSO/galaxy templates, or even linear combinations of the two. 
Moreover, point-like sources with X-ray-to-optical flux ratio 
$Log(f_{X}/f_{opt})<-1$ (i.e. in the region occupied mostly by Galactic 
stars; Stocke et al. 1991) are fitted with both QSO and stellar templates. 
Kitsionas et al. (2005) have demonstrated that this method is very efficient 
in identifying Galactic stars. In our sample there are in total 4 
$Log(f_{X}/f_{opt})<-1$ point-like sources which are best fitted by stellar 
templates. These sources are excluded from our final catalogue.

Given the fact that our sample is similar in nature to the sample 
discussed by Kitsionas et al. (2005), we adopt the same probabilities 
for obtaining reliable photometric redshifts. In particular, we assume 
that photometric redshift estimates for QSOs in our photometric 
sample have $\sim70$\% probability to be within 0.3 from the source's 
real redshift, whereas for red extended objects the probability for 
photometric redshift estimates to be within 0.15 from the source's real 
redshift is increased to $\sim75$\%. Our final sample, hereafter the 
{\xmm}/SDSS sample, comprises 26 sources without optical counterpart, 
45 sources spectroscopically classified, based on either SDSS or 
 NASA Extragalactic Database (NED) classification, 
as QSOs and 82 sources with photometric observations most probably 
associated with an AGN. 
In the case of the optically unidentified objects, we assume 
a mean redshift of 1.5 in order to estimate the source luminosity and
the column density. Although this is clearly an arbitrary assumption,  
the fact that SDSS did not detect these sources suggests that their 
redshift is likely higher than 1 . Indeed, galaxies with e.g. Mr=-21 
should be detected up to a redshift of 0.7 at the SDSS magnitude 
limit of r=22.5. Moreover the vast majority of these optically 
faint sources (22 out of 26) present $\log(f_{X}/f_{opt})>1$.
Previous work in both deep \chandra and shallower {\xmm} surveys 
(i.e. Alexander et al. 2002, Fiore et al. 2003) 
have shown that a considerable fraction of these sources
are type-II QSOs. 

In order to increase our statistics we have included in our {\xmm}/SDSS  
sample the data from HELLAS2XMM survey (Perola et al. 2004). There are 
44 optically identified, hard X-ray selected QSOs in the 2-8 keV flux 
limit of 3$\times 10^{-14}$ \funits. The conversion of their 2-10 keV 
intrinsic flux to our 2-8 keV band was made using an average photon 
index of 1.8.

\begin{figure}
   \centering
   \includegraphics[width=8.5cm]{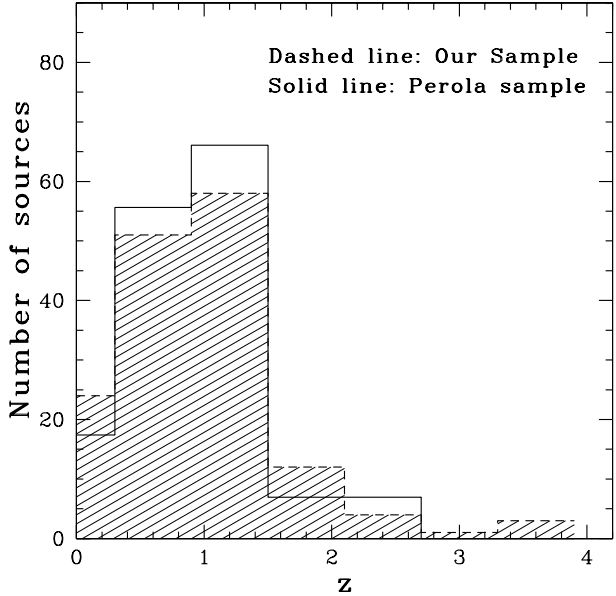}
   \caption{The redshift distribution of our {\xmm}/SDSS survey 
		sample comprising 153 sources (shaded histogram),
		including the 26 optically unidentified sources, 
		arbitrarily placed in the 1.0-1.5  redshift 
		bin (see text). We also plot the  normalized to our data
		redshift distribution of the 44 sources from the HELLAS2XMM 
		survey (Perola et al. 2004, open histogram).}
    \label{zdist}
    \end{figure}

In Fig. \ref{zdist} we plot the redshift distribution for the two
datasets. Hereafter we consider these combined observations as the 
{\xmm} sample. The final {\xmm} catalogue comprises 197 sources 
of which 171 have photometric or spectroscopic information and 
26 optically are unidentified sources to the flux limit of the 
SDSS survey. 

\section{THE X-RAY SPECTRAL ANALYSIS}

\subsection{\xmm/SDSS spectral analysis}

We explore the X-ray properties of the {\xmm}/SDSS sample using the 
{\sl XSPEC} v11.2 package to perform X-ray spectral fittings. 
For the sources with adequate count statistics (net source counts 
$>$ 100) we use $\chi^2$ statistic technique. The data are grouped 
to give a minimum of 15 counts per bin to ensure that Gaussian 
statistics apply. We adopt an absorbed power-law model and attempt 
to constrain the intrinsic absorption column density $N_H$ (i.e. having 
subtracted the Galactic absorption) and the power-law photon 
index $\Gamma$. 

For the sources with limited photon statistics (net counts $<$100) 
we use the C-statistic technique (Cash 1979) specifically developed 
to extract spectral information from data with low signal-to-noise 
ratio. In this case the data are grouped to give a minimum of 1 
count per bin to avoid zero count bins. 
We try to constrain the intrinsic column densities using an absorbed 
power-law model with $\Gamma$ fixed to 1.8. 

In both cases the spectral fittings are performed in the 0.3-8 keV 
energy band were the sensitivity of the {\xmm} detectors is the 
highest. The estimated errors correspond to the 90 per cent 
confidence level. In Table \ref{table2} we present the spectral 
fitting results for the 153 sources comprising our final {\xmm}/SDSS 
dataset. The spectral results for the 44 HELLAS2XMM survey 
sources have been obtained from Perola et al. (2004) and are not 
presented here. 
In columns 2 and 3 we list the source coordinates. In column 4 we 
give the best fit power-law photon index (when fixed to 1.8 
spectral fittings are performed using C-statistic).
In column 5 we present the best fit observed column density 
(uncorrected for redshift and  Galactic absorption) and in column 
6 the intrinsic 2-8 keV X-ray flux. Column 7 lists the source 
redshift and column 8 the value of the Galactic absorption in the 
direction of the pointing. In column 9 we present the intrinsic 
column density (i.e. corrected for redshift and Galactic absorption) 
and in column 10 the intrinsic hard (2-8 keV) X-ray luminosity using 
the K-correction appropriate for the best fit value of $\Gamma$. 
In the last column we have included an identification key in order to 
distinguish between data  with spectroscopic observations (id = 1), 
photometric observations (id = 2) or without optical identification 
(id = 3). 

The calculation of the intrinsic column density is based on the formula
${N_H}_{intr}=({N_H}_{obs}-{N_H}_{gal})\cdot(1+z)^{2.65}$. We have 
assumed a minimum value for the intrinsic column density, 
${N_H}_{intr}^{min}=({N_H}_{obs}^{min})\cdot(1+z)^{2.65}$, where
${N_H}_{obs}^{min}$=10$^{20}$ cm$^{-2}$ is approximately the 
minimum  column density that can be detected in the observed 0.3-8 keV 
energy range by the {\xmm} EPIC PN or MOS CCDs. Also in the cases 
where the spectral fitting analysis indicates a very flat spectrum
(i.e. $\Gamma<$1.0, see the sources \#50 and \#85 in Table \ref{table2}), 
 we assume that these sources are Compton thick and the 
$N_H$ value is arbitrarily set to 5$\times10^{24}$ cm$^{-2}$.

\begin{table*} 
\centering
\caption{The X-ray spectral fitting results for the 153 sources of our \xmm/SDSS sample.}
\label{table2}
\begin{tabular}{ccccccccccc}

\hline 
No & RA (J2000) & DEC (J2000) & $\Gamma^1$ & ${N_H}_{obs}^2$ & $F_{2-8 keV}^3$  & $z$ & ${N_H}_{gal}^4$& ${N_H}_{intr}^5$ & $Log(L_{2-8 keV})^6$ &  ID$^7$ \\
\hline
\hline
1   &  10.696  &   1.005 & 2.56$^{+0.54}_{-0.40}$ & $<$0.06 	           &     2.18&  1.00& 2.32&  0.06 & 44.03 & 2 \\ 
2   &  10.710  &   1.035 & 2.77$^{+0.28}_{-0.23}$ & $<$0.05 	           &     2.43&  0.60& 2.34&  0.03 & 43.56 & 2 \\
3   &  10.802  &   0.934 & 1.93$^{+0.13}_{-0.11}$ & $<$0.02 	 	   &     7.60&  0.94& 2.32&  0.06 & 44.51 & 2 \\
4   &  10.908  &   0.770 & 1.85$^{+0.31}_{-0.25}$ & $<$0.05 		   &     3.50&  2.13& 2.31&  0.21 & 44.99 & 2 \\
5   &  10.921  &   0.881 & 1.68$^{+0.36}_{-0.29}$ & 0.25$^{+0.16}_{-0.12}$ &     5.54&  4.40& 2.32&  20.10& 45.87 & 2 \\
6   &  10.922  &   0.935 & 1.92$^{+0.12}_{-0.10}$ & $<$0.02 	           &     8.71&  0.94& 2.32&  0.06 & 44.57 & 2 \\
7   &   10.959 &   0.963 & 2.10$^{+0.2}_{-0.16}$  & 0.14$^{+0.05}_{-0.04}$ &     17.8&  0.10& 2.34&  0.15 & 42.71 & 2 \\
8   &   11.056 &  0.861  & 2.31$^{+0.44}_{-0.32}$ & $<$0.06                &     4.66&  0.94& 2.34&  0.06 & 44.30 & 1 \\
9   &  28.011  &   1.079 & 1.80                   & $<$0.02                &     3.83&  1.05& 2.80&  0.07 & 44.32 & 2 \\
10  &   28.020 &   1.134 & 1.80                   & 0.30$^{+0.18}_{-0.14}$ &     4.21&  0.40& 2.80&  0.68 & 43.40 & 2 \\
11  &   28.126 &   0.950 & 1.80                   & 0.07$^{+0.02}_{-0.03}$ &     3.31&  0.35& 2.80&  0.11 & 43.17 & 2 \\
12  &   28.159 &   1.156 & 1.80                   & 1.08$^{+0.77}_{-0.38}$ &     4.19&  0.10& 2.79&  1.36 & 42.08 & 2 \\
13  &  28.179  &   1.205 & 1.80                   & 0.50$^{+1.08}_{-0.36}$ &     2.36&  1.40& 2.79&  4.88 & 44.41 & 2 \\
14  &  28.190  &   1.184 & 1.80                   & $<$0.06                &     1.59&  1.51& 2.79&  0.11 & 44.30 & 2 \\
15  &   28.244 &  1.085  & 2.59$^{+0.55}_{-0.36}$ & $<$0.10		   &     1.78&  0.65& 2.80&  0.04 & 43.50 & 1 \\
16  &   28.280 &   1.098 & 1.65$^{+0.35}_{-0.33}$ & $<$0.075		   &     7.16&  0.20& 2.80&  0.01 & 42.95 & 2 \\
17  &   29.820 &   0.504 & 2.10$^{+0.67}_{-0.38}$ & $<$0.12		   &     10.8&  0.35& 2.60&  0.02 & 43.67 & 2 \\
18  &   29.990 &  0.553  & 2.09$^{+0.16}_{-0.15}$ & $<$0.02		   &     26.9&  0.31& 2.62&  0.02 & 43.96 & 1 \\ 
19  &   30.105 &   0.488 & 1.80                   & $<$0.41		   &     4.82&  0.30& 2.60&  0.02 & 43.18 & 2 \\
20  &   30.120 &  0.480  & 2.34$^{+0.67}_{-0.28}$ & $<$0.06		   &     6.87&  0.17& 2.60&  0.02 & 42.81 & 1 \\
21  &   40.090 &  -8.361 & 1.80                   & $<0.02$     	   &     4.15&  1.30& 3.11&  0.09 & 44.57 & 2 \\
22  &  40.106  &  -8.408 & 2.15$^{+0.17}_{-0.18}$ & 0.15$^{+0.05}_{-0.05}$ &     31.3&     -& 3.12&  1.42 & 45.59 & 3 \\
23  &  40.169  &  -8.295 & 1.80                   & 0.15$^{+0.22}_{-0.12}$ &     2.16&     -& 3.09&  1.47 & 44.43 & 3 \\
24  &   40.327 &  -8.428 & 1.80                   & $<$0.04		   &     3.69&  1.43& 3.09&  0.10 & 44.62 & 2 \\
25  &  44.025  &   0.276 & 1.80                   &  $<$0.02		   &     4.95&  0.60& 6.50&  0.03 & 43.87 & 2 \\
26  &   44.113 &   0.126 & 1.80                   & $<$0.09		   &     3.48&  0.34& 6.50&  0.44 & 43.17 & 2 \\
27  &  44.158  &   0.234 & 1.80                   & 2.05$^{+1.12}_{-1.25}$ &     4.11&     -& 6.52&  23.01& 44.71 & 3 \\
28  &   44.188 &   0.008 & 1.80                   & $<$0.05		   &     6.59&  0.50& 6.51&  0.07 & 43.81 & 2 \\
29  &  44.211  &   0.226 & 1.80                   &  $<$0.032		   &     3.79&  1.00& 6.66&  1.88 & 44.27 & 2 \\
30  &  45.518  &   0.274 & 1.80$^{+0.15}_{-0.22}$ & 0.04$^{+0.05}_{-0.04}$ &     5.24&  4.28& 7.17&  1.97 & 45.82 & 2 \\
31  &  45.527  &  -0.022 & 1.88$^{+0.11}_{-0.09}$ & 0.07$^{+0.02}_{-0.02}$ &     17.1&  0.64& 7.15&  0.19 & 44.48 & 1 \\
32  &  45.563  &  -0.059 & 1.80                   & 0.35$^{+0.07}_{-0.08}$ &     2.32&  4.53& 7.15&  30.67& 45.52 & 2 \\
33  &  45.780  &   0.171 & 2.17$^{+0.15}_{-0.13}$ & 0.07$^{+0.02}_{-0.03}$ &     3.53&  1.02& 7.08&  0.36 & 44.26 & 2 \\
34  &  54.475  &   0.493 & 1.80                   & $<$0.04		   &     4.26&  2.02& 8.22&  0.19 & 45.02 & 2 \\
35  &   54.542 &  0.390  & 2.35$^{+0.54}_{-0.39}$ & $<$0.15		   &     2.74&  1.12& 8.12&  0.18 & 44.25 & 1 \\
36  &  54.562  &   0.490 & 2.20$^{+0.20}_{-0.25}$ &$<$0.2		   &     9.13&     -& 8.21&  0.11 & 45.06 & 3 \\
37  &   54.577 &  0.198  & 1.80                   & $<$0.05		   &     3.11&  1.58& 7.92&  0.12 & 44.65 & 1 \\
38  &  54.616  &   0.420 & 1.80                   & 0.06$^{+0.02}_{-0.03}$ &     2.61&  1.30& 8.15&  0.36 & 44.37 & 2 \\
39  &   54.656 &   0.346 & 1.80                   & 2.80$^{+4.20}_{-1.60}$ &     3.40&  0.70& 8.10&  11.33&  43.86&  2 \\ 
40  &  54.677  &   0.357 & 1.80                   & 0.09$^{+0.04}_{-0.04}$ &     2.44&  1.52& 8.11&  0.81 & 44.48 & 2 \\
41  &  59.203  &   1.359 & 2.20$^{+0.55}_{-0.38}$ & 0.36$^{+0.18}_{-0.14}$ &     7.13&     -& 13.3&  3.85 & 44.95 & 3 \\ 
42  &  59.243  &   1.349 & 1.80                   & 0.09$^{+0.06}_{-0.05}$ &     3.22&     -& 13.3&  0.84 & 44.60 & 3 \\  
43  &  59.282  &   1.260 & 2.35$^{+0.29}_{-0.33}$ & 0.27$^{+0.18}_{-0.13}$ &     3.00&     -& 13.2&  2.83 & 44.58 & 3 \\
44  &  59.322  &   1.295 & 1.85$^{+0.16}_{-0.14}$ & $<$0.04		   &     5.87&     -& 13.3&  0.11 & 44.87 & 3 \\
45  &   59.480 &   1.239 & 1.61$^{+0.53}_{-0.38}$ & $<$0.22		   &     5.42&  1.51& 13.3&  0.74 & 44.84 & 2 \\
46  & 127.611  &  52.694 & 2.04$^{+0.09}_{-0.11}$ & 0.09$^{+0.02}_{-0.03}$ &     5.86&     -& 3.96&  0.73 & 44.87 & 3 \\
47  &  127.611 &  52.767 & 1.89$^{+0.15}_{-0.15}$ & 0.06$^{+0.04}_{-0.04}$ &     4.57&  0.60& 3.96&  0.15 & 43.83 & 2 \\
48  & 127.707  &  52.819 & 1.76$^{+0.10}_{-0.09}$ & 0.03$^{+0.02}_{-0.02}$ &     9.19&  1.40& 3.89&  0.10 & 44.99 & 2 \\
49  &  127.785 &  52.644 & 1.86$^{+0.14}_{-0.17}$ & 0.38$^{+0.16}_{-0.11}$ &     4.13&  0.20& 3.87&  0.58 & 42.72 & 2 \\
50  & 127.821  &  52.588 & 0.25$^{+0.45}_{-0.25}$ &$>$500 		   &     6.87&     -& 3.85&  500  & 44.93 & 3 \\

\hline
\end{tabular} 								
\end{table*}     

\begin{table*} 
\centering
\begin{tabular}{ccccccccccc}

\hline 
No & RA(J2000) & DEC(2000) & $\Gamma^1$ & ${N_H}_{obs}^2$ & $F_{2-8 keV}^3$ & $z$ & ${N_H}_{gal}^4$& ${N_H}_{intr}^5$ & $L_{2-8 keV}^6$ &  ID$^7$ \\
\hline
\hline

51  &  127.822 &  52.815 & 1.81$^{+0.07}_{-0.11}$ & $<$0.05		   &     3.96&  0.70& 3.87&  0.05 & 43.93 & 2 \\
52  &  127.912 & 52.701  & 1.80                   &18.33$^{+2.17}_{-3.13}$ &     17.0&  0.05& 3.84&  21.30&  42.21&  1 \\
53  &  128.020 &  52.922 & 1.80                   & 3.30$^{+0.67}_{-0.88}$ &     5.76&  0.99& 3.84&  20.53&  44.45&  2 \\
54  & 128.076  &  52.623 & 2.05$^{+0.12}_{-0.17}$ & $<$0.07		   &     2.63&  1.22& 3.79&  0.08 & 44.31 & 2 \\
55  &  139.188 &  51.695 & 1.72$^{+0.17}_{-0.14}$ & $<$0.026		   &     12.4&  0.50& 1.47&  0.03 & 44.09 & 2 \\
56  &  139.516 &  51.687 & 1.80                   & 0.16$^{+0.03}_{-0.05}$ &     3.13&  0.10& 1.45&  0.18 & 41.95 & 2 \\
57  &  139.821 &  51.691 & 1.80                   & 0.33$^{+0.47}_{-0.22}$ &     2.46&  0.80& 1.41&  1.47 & 43.86 & 2 \\
58  &  143.417 &  55.111 & 2.64$^{+0.56}_{-0.29}$ & $<$0.08		   &     1.50&  0.90& 1.89&  0.05 & 43.76 & 2 \\
59  &  143.450 &  55.312 & 1.65$^{+0.08}_{-0.09}$ & 0.10$^{+0.03}_{-0.02}$ &     17.7&  0.60& 1.89&  0.28 & 44.42 & 2 \\
60  &  143.497 &  55.263 & 1.82$^{+0.11}_{-0.16}$ & $<$0.05		   &     4.60&  1.70& 1.88&  0.14 & 44.88 & 2 \\
61  &  143.650 &  61.270 & 1.60$^{+0.36}_{-0.25}$ & $<$0.02		   &     3.89&  1.90& 2.69&  0.15 & 44.87 & 2 \\
62  &  143.744 & 61.209  & 1.25$^{+0.55}_{-0.4}$  & 1.40$^{+1.22}_{-1.10}$ &     15.2&  0.25& 2.65&  2.47 & 43.48 & 1 \\
63  &  143.889 & 61.461  & 1.80                   & 2.14$^{+0.32}_{-0.60}$ &     4.56&  0.47& 2.72&  5.95 & 43.61 & 1 \\
64  &  143.898 &  61.322 & 2.46$^{+0.21}_{-0.14}$ & 0.04$^{+0.03}_{-0.02}$ &     3.85&  0.50& 2.67&  0.05 & 43.58 & 2 \\	
65  & 144.030  &  61.546 & 2.62$^{+0.28}_{-0.32}$ & 0.16$^{+0.06}_{-0.09}$ &     2.50&     -& 2.76&  1.54 & 44.50 & 3 \\
66  &  144.035 &  61.507 & 1.78$^{+0.41}_{-0.23}$ & $<$0.072		   &     3.52&  0.60& 2.73&  0.03 & 43.72 & 2 \\
67  & 144.266  &  61.266 & 1.80                   & 0.29$^{+0.68}_{-0.20}$ &     0.72&     -& 2.51&  3.06 & 43.95 & 3 \\
68  &  144.270 &  61.440 & 2.17$^{+0.54}_{-0.37}$ & $<$0.06		   &     2.55&  1.43& 2.57&  0.38 & 44.46 & 2 \\
69  & 144.373  &  61.432 & 1.80                   & 1.74$^{+1.03}_{-0.71}$ &     3.62&     -& 2.56&  19.50& 44.66 & 3 \\
70  &  187.651 &  64.425 & 1.80                   & $<$0.05		   &     3.47&  0.60& 2.05&  0.03 & 43.72 & 2 \\
71  & 187.793  &  64.313 & 1.80                   & 0.04$^{+0.01}_{-0.02}$ &     3.58&  0.60& 2.00&  0.06 & 43.73 & 2 \\
72  & 188.077  &  64.052 & 1.89$^{+0.25}_{-0.27}$ & 0.05$^{+0.06}_{-0.05}$ &     5.64&  2.20& 1.94&  0.61 & 45.22 & 2 \\
73  &  191.103 &  -0.410 & 1.73$^{+0.09}_{-0.07}$ & $<$0.02		   &     8.75&  0.40& 1.76&  0.02 & 43.71 & 2 \\
74  &  191.122 &  -0.571 & 1.66$^{+0.52}_{-0.37}$ & 0.38$^{+0.36}_{-0.26}$ &     7.70&  1.01& 1.77&  2.26 & 44.58 & 2 \\
75  & 191.124  &  -0.413 & 1.44$^{+0.27}_{-0.48}$ & 0.17$^{+0.19}_{-0.15}$ &     5.45&     -& 1.76&  1.70 & 44.83 & 3 \\
76  & 191.170  &  -0.420 & 1.95$^{+0.14}_{-0.16}$ & 0.19$^{+0.04}_{-0.06}$ &     4.38&     -& 1.75&  1.92 & 44.74 & 3 \\
77  & 191.224  &  -0.356 & 1.92$^{+0.09}_{-0.12}$ & 0.08$^{+0.03}_{-0.03}$ &     4.01&  1.10& 1.74&  0.44 & 44.39 & 2 \\
78  &  191.240 & -0.271  & 1.93$^{+0.27}_{-0.19}$ & $<$1.83		   &     3.98&  0.12& 1.73&  0.07 & 42.22 & 1 \\ 
79  & 191.250  &  -0.355 & 1.82$^{+0.15}_{-0.05}$ &  $<$0.04		   &     3.09&  1.30& 1.74&  0.09 & 44.45 & 2 \\
80  &  191.317 & -0.316  & 1.88$^{+0.06}_{-0.09}$ & $<$0.03		   &     6.61&  1.58& 1.73&  0.12 & 44.97 & 1 \\
81  &  191.421 & -0.463  & 1.80$^{+0.05}_{-0.07}$ & 0.03$^{+0.02}_{-0.02}$ &     11.1&  1.69& 1.73&  0.18 & 45.27 & 1 \\
82  &  191.490 &  -0.510 & 1.66$^{+0.26}_{-0.17}$ & $<$0.05		   &     3.18&  0.40& 1.73&  0.02 & 43.27 & 2 \\
83  &  195.740 & 67.501  & 1.98$^{+0.31}_{-0.19}$ & $<$0.04		   &     3.01&  1.84& 1.85&  0.16 & 44.78 & 1 \\
84  &  195.903 &  67.507 & 1.63$^{+0.36}_{-0.17}$ & $<$0.06		   &     15.9&  0.30& 1.85&  0.02 & 43.69 & 2 \\
85  &  195.996 &  67.509 & 0.75$^{+0.46}_{-0.31}$ & $<$0.02		   &     11.5&  0.50& 1.85&  500  & 44.05 & 2 \\
86  &  196.225 & 67.501  & 2.38$^{+0.16}_{-0.12}$ & $<$0.02		   &     2.93&  0.54& 1.85&  0.03 & 43.54 & 1 \\ 
87  &  196.459 &  67.655 & 1.63$^{+0.26}_{-0.17}$ & $<$0.07		   &     16.0&  0.50& 1.85&  0.03 & 44.20 & 2 \\
88  & 205.150  &  -0.446 & 1.80                   & 3.88$^{+8.78}_{-3.04}$ &     4.81&     -& 1.93&  43.76& 44.78 & 3 \\       
88  &  205.161 &  0.321  & 2.64$^{+0.74}_{-0.54}$ & $<$0.20		   &     4.22&  0.50& 1.84&  0.32 & 43.62 & 2 \\
90  &  205.188 & -0.400  & 1.80                   & 0.46$^{+0.30}_{-0.18}$ &     5.12&  0.50& 1.91&  1.29 & 43.70 & 2 \\   
91  & 205.209  &   0.265 & 1.80                   & 0.48$^{+0.18}_{-0.39}$ &     4.15&     -& 1.84&  5.21 & 44.72 & 3 \\
92  &  205.325 & -0.389  & 1.70$^{+0.30}_{-0.20}$ & $<$0.02		   &     9.81&  0.42& 1.93&  0.03 & 43.82 & 1 \\
93  &  205.339 & -0.230  & 1.80                   & $<$0.02		   &     5.43&  0.73& 1.87&  0.04 & 44.11 & 1 \\ 
94  &  205.363 &  0.237  & 1.80                   & $<$0.02		   &     4.18&  1.69& 1.85&  0.14 & 44.84 & 1 \\
95  &  205.368 & -0.522  & 1.80                   & 1.78$^{+0.22}_{-0.61}$ &     16.4&  0.60& 2.01&  6.12 & 44.39 & 2 \\
96  &  205.418 &  0.262  & 1.80                   & 0.05$^{+0.05}_{-0.03}$ &     3.23&  0.25& 1.85&  0.06 & 42.84 & 1 \\
97  &  205.428 &  0.210  & 1.79$^{+0.06}_{-0.06}$ & $<$0.03		   &     11.0&  0.79& 1.85&  0.05 & 44.54 & 1 \\
98  &  205.487 &  0.502  & 1.80                   & $<$0.02		   &     2.19&  1.23& 1.86&  0.08 & 44.25 & 1 \\
99  &  205.550 &  0.497  & 1.80                   & 0.19$^{+0.38}_{-0.16}$ &     1.66&  0.57& 1.86&  0.56 & 43.35 & 1 \\
100 &  205.643 &   0.539 & 1.80                   & $<$22 		   &     3.80&  2.01& 1.87&  0.18 & 44.96 & 2 \\
101 & 205.678  &   0.539 & 1.80                   & $<$0.07		   &     4.15&     -& 1.88&  0.62 & 44.72 & 3 \\
102 &  205.692 & -0.595  & 1.80                   & $<$0.04		   &     7.92&  0.79& 2.10&  0.05 & 44.35 & 1 \\ 
103 &  205.731 &  0.110  & 1.80                   & $<$0.02		   &     6.68&  0.44& 1.89&  0.03 & 43.69 & 1 \\
\hline
\end{tabular} 								
\end{table*}     
\begin{table*}
\centering
\begin{tabular}{ccccccccccc}

\hline 
No & RA(J2000) & DEC(J2000) & $\Gamma^1$ & ${N_H}_{obs}^2$ & $F_{2-8 keV}^3$ & $z$ & ${N_H}_{gal}^4$& ${N_H}_{intr}^5$ & $L_{2-8 keV}^6$ &  ID$^7$ \\
\hline
\hline
104 &  205.735 &  0.015  & 1.93$^{+0.62}_{-0.4}$  & 0.03$^{+0.26}_{-0.03}$ &     10.9&  0.80& 1.91&  0.07 & 44.51 & 1 \\
105 &  205.756 &   0.442 & 1.80                   & 0.06$^{+0.03}_{-0.03}$ &     5.13&  1.51& 1.89&  0.45 & 44.81 & 2 \\
106 &  205.849 &  0.205  & 1.80                   & $<$0.02		   &     10.6&  0.87& 1.90&  0.05 & 44.58 & 1 \\ 
107 &  205.871 &  0.026  & 1.80                   & $<$0.15		   &     4.46&  2.35& 1.94&  0.25 & 45.19 & 1 \\
108 &  205.881 &   0.413 & 1.80                   & $<$0.24		   &     1.30&  1.20& 1.90&  0.08 & 43.99 & 2 \\
109 &  205.886 & -0.034  & 1.80                   & $<$0.13		   &     3.02&  1.60& 1.96&  0.12 & 44.64 & 1 \\
110 &  205.928 &  0.300  & 1.80                   & 0.15$^{+0.16}_{-0.09}$ &     5.63&  1.11& 1.91&  0.95 & 44.55 & 1 \\
111 &  205.948 &  0.339  & 1.80                   & $<$0.04		   &     12.3&  0.24& 1.91&  0.02 & 43.36 & 1 \\
112 & 205.949  &  -0.033 & 1.80                   & 0.39$^{+0.94}_{-0.39}$ &     2.66&     -& 1.97&  4.19 & 44.52 & 3 \\
113 &  205.963 &  0.077  & 1.80                   & $<$0.02		   &     7.45&  0.07& 1.95&  0.01 & 42.05 & 1 \\
114 & 205.968  &  -0.075 & 1.80                   & 1.13$^{+0.67}_{-0.67}$ &     9.73&     -& 1.99&  12.58& 45.09 & 3 \\
115 &  206.083 &  0.071  & 1.80                   & 0.35$^{+0.15}_{-0.15}$ &     5.72&  0.30& 1.97&  0.66 & 43.25 & 2 \\
116 &  206.083 & -0.519  & 1.80                   & $<$0.02		   &     2.42&  0.68& 2.15&  0.04 & 43.69 & 1 \\
117 &  206.092 & -0.572  & 1.80                   & 0.27$^{+0.10}_{-0.07}$ &     2.45&  0.22& 2.17&  0.42 & 42.58 & 1 \\
118 &  206.102 & -0.218  & 1.80                   & $<$0.03		   &     4.48&  1.11& 2.06&  0.07 & 44.45 & 1 \\ 
119 &  206.105 & -0.307  & 1.80                   & 0.16$^{+0.10}_{-0.10}$ &     2.55&  1.97& 2.09&  2.51 & 44.77 & 1 \\
120 &  206.151 &  0.556  & 1.80                   & $<$0.02		   &     3.02&  1.43& 1.94&  0.10 & 44.53 & 1 \\
121 & 206.161  &  -0.183 & 1.80                   & 2.82$^{+3.51}_{-1.70}$ &     8.03&     -& 2.07&  31.74& 45.00 & 3 \\
122 &  206.212 &   0.279 & 1.80                   & 0.12$^{+0.03}_{-0.07}$ &     3.82&  1.43& 1.97&  1.05 & 44.63 & 2 \\
123 &  206.220 &  0.089  & 2.34$^{+0.11}_{-0.11}$ & $<$0.02		   &     28.0&  0.08& 2.02&  0.01 & 42.78 & 1 \\
124 &  206.241 & -0.600  & 1.80                   &11.60$^{+18.40}_{-4.60}$&	13.2 &  0.46& 2.21& 31.86 & 44.04 & 1 \\
125 &  206.243 &  0.272  & 1.80                   & $<$0.03		   &     4.82&  0.15& 1.98&  0.01 & 42.48 & 1 \\
126 &  206.248 & -0.266  & 1.90$^{+0.41}_{-0.34}$ & $<$0.02		   &     21.1&  0.24& 2.12&  0.02 & 43.62 & 1 \\ 
127 &  206.283 & -0.091  & 1.80                   & $<$0.02		   &     8.51&  0.73& 2.09&  0.04 & 44.30 & 1 \\ 
128 &  206.290 &  0.347  & 1.80                   & 0.27$^{+0.19}_{-0.13}$ &     5.89&  0.50& 1.98&  0.73 & 43.76 & 2 \\
129 & 207.012  &  60.109 & 1.80                   & 2.38$^{+2.50}_{-1.43}$ &     2.67&     -& 1.82&  26.75& 44.52 & 3 \\
130 &  207.248 &  60.250 & 1.95$^{+0.17}_{-0.11}$ & $<$0.03		   &     6.61&  0.50& 1.81&  0.03 & 43.81 & 2 \\
131 &  207.367 &  60.104 & 1.89$^{+0.26}_{-0.15}$ & $<$0.06		   &     4.53&  0.10& 1.80&  0.01 & 42.11 & 2 \\
132 & 207.492  &  60.330 & 1.80                   & 1.64$^{+1.70}_{-0.34}$ &     5.91&     -& 1.80&  18.34& 44.87 & 3 \\
133 &  235.760 & 54.153  & 1.80                   & 0.03$^{+0.03}_{-0.01}$ &     4.38&  0.61& 1.25&  0.04 & 43.84 & 2 \\
134 &  235.967 &  54.001 & 2.35$^{+0.25}_{-0.17}$ & 0.03$^{+0.04}_{-0.02}$ &     3.30&  1.48& 1.25&  0.13 & 44.60 & 2 \\
135 &  235.998 &  53.984 & 2.30$^{+0.14}_{-0.11}$ & 0.12$^{+0.05}_{-0.05}$ &     5.57&  2.37& 1.25&  2.50 & 45.29 & 1 \\
136 & 236.007  &  54.127 & 1.80                   & $<$0.02		   &     2.24&     -& 1.24&  0.11 & 44.45 & 3 \\
137 &  236.101 &  53.929 & 1.49$^{+0.40}_{-0.17}$ & $<$0.10		   &     9.68&  0.70& 1.25&  0.04 & 44.32 & 2 \\
138 &  236.147 & 54.094  & 1.95$^{+0.47}_{-0.32}$ &  $<$0.07		   &     3.19&  1.00& 1.24&  0.06 & 44.19 & 2 \\
139 &  255.174 &  64.216 & 1.84$^{+0.76}_{-0.3}$  &  $<$0.25		   &     21.8&  0.20& 2.67&  0.02 & 43.44 & 2 \\
140 &  255.196 &  64.384 & 1.80                   & 2.20$^{+3.50}_{-1.12}$ &     16.9&  0.60& 2.69&  7.57 & 44.40 & 2 \\
141 &  255.252 & 64.202  & 1.96$^{+0.54}_{-0.3}$  &  $<$0.25		   &     7.49&  2.73& 2.66&  0.33 & 45.56 & 1 \\
142 &  255.349 & 64.236  & 2.12$^{+0.53}_{-0.36}$ & $<$0.16		   &     10.2&  0.45& 2.65&  0.06 & 43.91 & 1 \\ 
143 &  354.221 &   0.347 & 1.80                   & $<$0.06		   &     3.02&  0.94& 3.85&  0.06 & 44.11 & 2 \\
144 & 354.312  &   0.376 & 1.80                   & 2.52$^{+2.48}_{-1.18}$ &     4.40&     -& 3.83&  28.34& 44.74 & 3 \\
145 &  354.382 &   0.433 & 1.80                   & $<$0.03		   &     4.14&  0.30& 3.82&  0.02 & 43.11 & 2 \\
146 &  354.408 &   0.268 & 1.80                   & 2.60$^{+0.38}_{-0.52}$ &     9.89&  0.40& 3.82&  6.29 & 43.77 & 2 \\
147 &  354.547 &  0.345  & 1.82$^{+0.23}_{-0.19}$ & $<$0.045		   &     8.15&  0.28& 3.79&  0.02 & 43.33 & 1 \\ 
148 & 354.582  &   0.186 & 1.80                   & 0.70$^{+0.30}_{-0.25}$ &     5.37&     -& 3.80&  7.70 & 44.83 & 3 \\
149 &  354.630 &   0.356 & 1.80                   & $<$0.48		   &     0.62&  0.62& 3.78&  0.04 & 43.00 & 2 \\
150 &  358.456 & -10.523 & 1.80                   & $<$0.03		   &     1.68&  0.70& 2.92&  0.04 & 43.56 & 2 \\
151 &  358.466 & -10.257 & 1.80                   & 0.21$^{+0.07}_{-0.07}$ &     4.56&  0.30& 2.86&  0.38 & 43.15 & 2 \\
152 &  358.540 & -10.359 & 1.71$^{+0.16}_{-0.13}$ & $<$0.03		   &     6.40&  0.55& 2.88&  0.03 & 43.89 & 2 \\
153 &  358.643 & -10.268 & 1.53$^{+0.22}_{-0.23}$ & 0.05$^{+0.11}_{-0.05}$ &     9.89&  1.36& 2.87&  0.30 & 44.99 & 2 \\
\hline

\end{tabular} 

\begin{list}{}{}
\item$^1$When the photon index is fixed to 1.8 the spectral fittings are 
performed using the C-statistic technique. \\
\item$^2$Observed, best fit $N_H$ value in units of $10^{22}$ cm$^{-2}$ \\
\item$^3$Intrinsic 2-8 keV flux in units
 of 10$^{-14}$ \funits.  \\
\item$^4$Galactic column density in the directions of the observed field 
in units of 10$^{20}$ cm$^{-2}$ \\
\item$^5$Intrinsic rest frame column density in units of $10^{22}$ cm$^{-2}$ \\
\item$^6$Logarithm of intrinsic 2-8 keV luminosity in units of \lunits.\\
\item$^7$The ID values (ID= 1, 2 or 3) indicate spectroscopic, 
photometric or the absence of an optical counterpart respectively. \\
\end{list}
\end{table*}

\subsection{The CDF-S spectral analysis}{\label{scdf}}

\begin{figure}
   \centering
   \includegraphics[width=8.5cm]{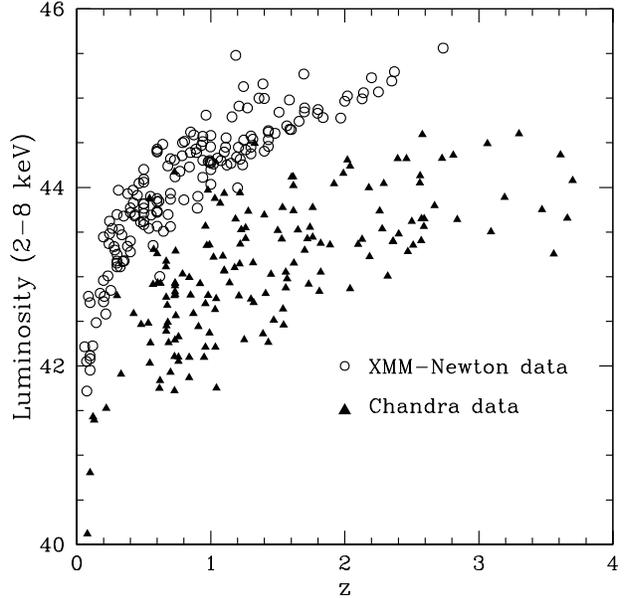}
   \caption{The 2-8 keV intrinsic luminosity  versus redshift plot for the 
		{\xmm} (open circles) and the {\chandra} (filled triangles) 
		sources.}
   \label{lz}
   \end{figure}

In order to cover a much broader flux range and therefore to examine the  
widest possible luminosity and redshift range we also include in 
our analysis the data from the \chandra
Deep Field South (CDF-S) (Giacconi et al. 2002).
In Fig. \ref{lz} we plot the intrinsic 2-8 keV luminosity versus 
redshift for the \chandra and the {\xmm} samples. 
The median luminosity for the {\chandra} dataset is 
$1.5^{+4.5}_{-1.1}\times10^{43}$ {\lunits}, one 
order of magnitude lower than that of the 
{\xmm} data ($2.1^{+3.0}_{-1.5}\times10^{44}$ \lunits)
due to the significant difference in the flux limit of the 
observations.

We estimate the $N_H$ using X-ray spectral 
fittings for all 162 CDF-S sources. 
The spectral files and the auxiliary files were produced 
from the merged CDF-S event file using the CIAO v.3.2 software, which 
also corrects the auxillary files for the degradation in 
the ACIS Quantum Efficiency due to molecular contamination.
We consider only the data in the 0.3-8 keV energy range.
We adopt a radius of 6 arcsec  for the source count extraction and 
a 10 times larger area for the 
background estimation. We use the {\sl DMARFADD} task of CIAO to 
add the ARF files from each observation separately to create 
a single output file for each source. Similarly the mean   
RMF file for each source is extracted using the {\sl ADDRMF} 
task of {\sl FTOOLS}. 

The X-ray spectral fittings are performed using the {\sl XSPEC} 
v11.2 package. We use the C-statistic technique (Cash 1979). 
The data are grouped to give a minimum of 1 count per bin to 
avoid zero count bins. We try to constrain the intrinsic column 
densities (i.e. having subtracted the Galactic absorption of 
8$\times10^{19}$ cm$^{-2}$, Dickey \& Lockman 1990) using an 
absorbed power-law model with $\Gamma$ fixed to 1.8. 

\section{RESULTS}

\subsection{The $N_H$ distribution}{\label{snh}}

\begin{figure}
   \centering
   \includegraphics[width=8.5cm]{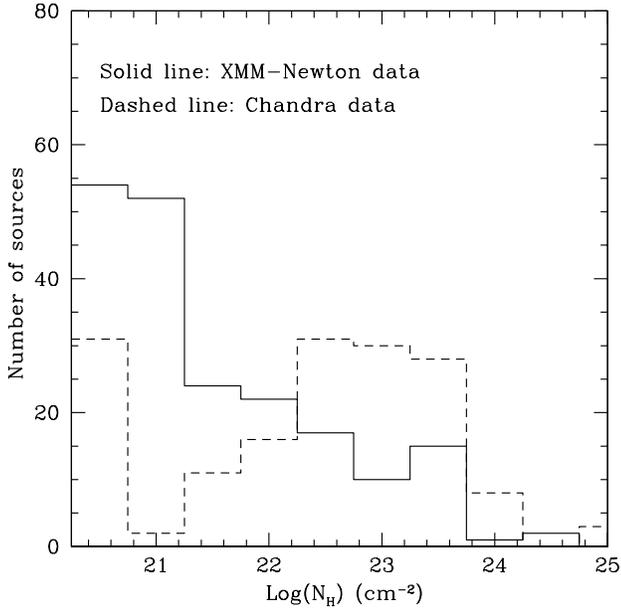}
   \caption{The $N_H$ distribution of the sources in the combined
		{\xmm} sample (solid line) compared
		to that derived from the CDF-S observations
		(dashed line).}
    \label{nhdist}
    \end{figure}

In Fig. \ref{nhdist} we plot the $N_H$ distribution for 
the {\xmm} and {\chandra} samples. The dashed line 
represents the $N_H$ distribution for the 162 sources in 
the CDF-S survey and the solid line shows the same 
distribution for the 171 optically identified sources 
in the combined {\xmm} sample.
The median fluxes are 
$2.0^{+2.9}_{-0.9}\times 10^{-15}$ \funits 
and $4.6^{+3.8}_{-1.1}\times 10^{-14}$ \funits 
and the median redshifts $1.0^{+0.7}_{-0.3}$ and
$0.8^{+0.5}_{-0.4}$ for the \chandra and the {\xmm} samples 
respectively. When we consider the 26 optically undetected 
sources at z=1.5 the median redshift
for the {\xmm} data becomes $0.9^{+0.5}_{-0.4}$. 
The errors in the median values correspond to the 
upper and lower probability quartiles.

Fig. \ref{nhdist} demonstrates that there is  a large difference between 
the $N_H$ distributions calculated from the {\xmm} and the 
{\chandra} data.
This difference could arise as the \chandra sample probes much deeper 
fluxes. It is well established (e.g. Alexander et al. 2003) that there 
is a strong correlation between the average hardness ratio and the 
flux in the sense that at fainter fluxes we probe harder sources.
On the other hand, the {\chandra} observations reveal 
intrinsically less luminous sources (by approximately an 
order of magnitude) (see Fig. \ref{lz}). Therefore it is 
possible that the difference between the absorption observed 
in the {\xmm} and {\chandra} samples may also be explained 
assuming that sources with lower luminosity present on average 
larger amounts of absorption. Next, we will attempt to 
disentangle between these two possibilities. 
  
\subsection{$N_H$--Luminosity dependence}

We use the data presented above to diagnose 
whether the strength of the photoelectric absorption depends 
on the intrinsic luminosity. The simplest way to detect such a 
possible correlation is to test whether there is a significant 
decrease in the fraction of obscured objects as the intrinsic 
luminosity increases. 
This technique is affected by a selection bias. We observe the 
unobscured sources within a larger volume compared to the obscured 
ones, as the observed luminosity of the latter decreases
substantially due to photoelectric absorption.
To account for this effect we calculate the fraction of absorbed 
sources using the $1/V_{m}$ method (see Page \& Carrera 2000). 
For each source of a given observed 2-8 keV luminosity $L_X$ 
we calculate the maximum available volume using the formula:

\begin{center}
$\displaystyle V_{m}=\int_{0}^{z_{max}} \Omega(f)\frac{dV}{dz}dz$
\end{center}

\noindent
where $\Omega(f)$ is the value of the sensitivity curve
at a given flux, corresponding to a source at a redshift $z$ 
with observed luminosity $L_X$ and $z_{max}$
the maximum redshift at which the source can be observed
at the flux limit of the survey.
The fraction of the obscured objects at a given luminosity bin 
is then calculated using the formula:

\begin{center}
$\displaystyle Fraction=\frac{\displaystyle\sum^{N_1}_{i=1}\frac{1}
{V_{m}(i)}}{\displaystyle\sum^{N}_{i=1}\frac{1}{V_{m}(i)}}$
\end{center}

\noindent
where N$_1$ is the number of the obscured sources 
($N_H>10^{22}$ cm$^{-2}$)
and N is the total number of the sources in each 
luminosity bin. The corresponding errors in 1/V$_{m}$ 
are approximated by:

\begin{center}
$\delta(1/V{_m})=\displaystyle\sum^{N_1}_{i=1}\frac{1}{V_{m}^{2}(i)}$
\end{center}

\noindent
while the errors in the fraction are estimated using the error 
propagation formula.

\noindent
We apply this correction in both {\xmm} and \chandra 
observations. The area curve of our {\xmm} survey is constant up to 
the flux limit of $3\times10^{-14}$ \funits (see Georgakakis 
et al. 2006a), while the area curve for the CDF-S has been adapted  
from Giacconi et al. (2002). 

\begin{figure}
   \centering
   \includegraphics[height=6.8cm]{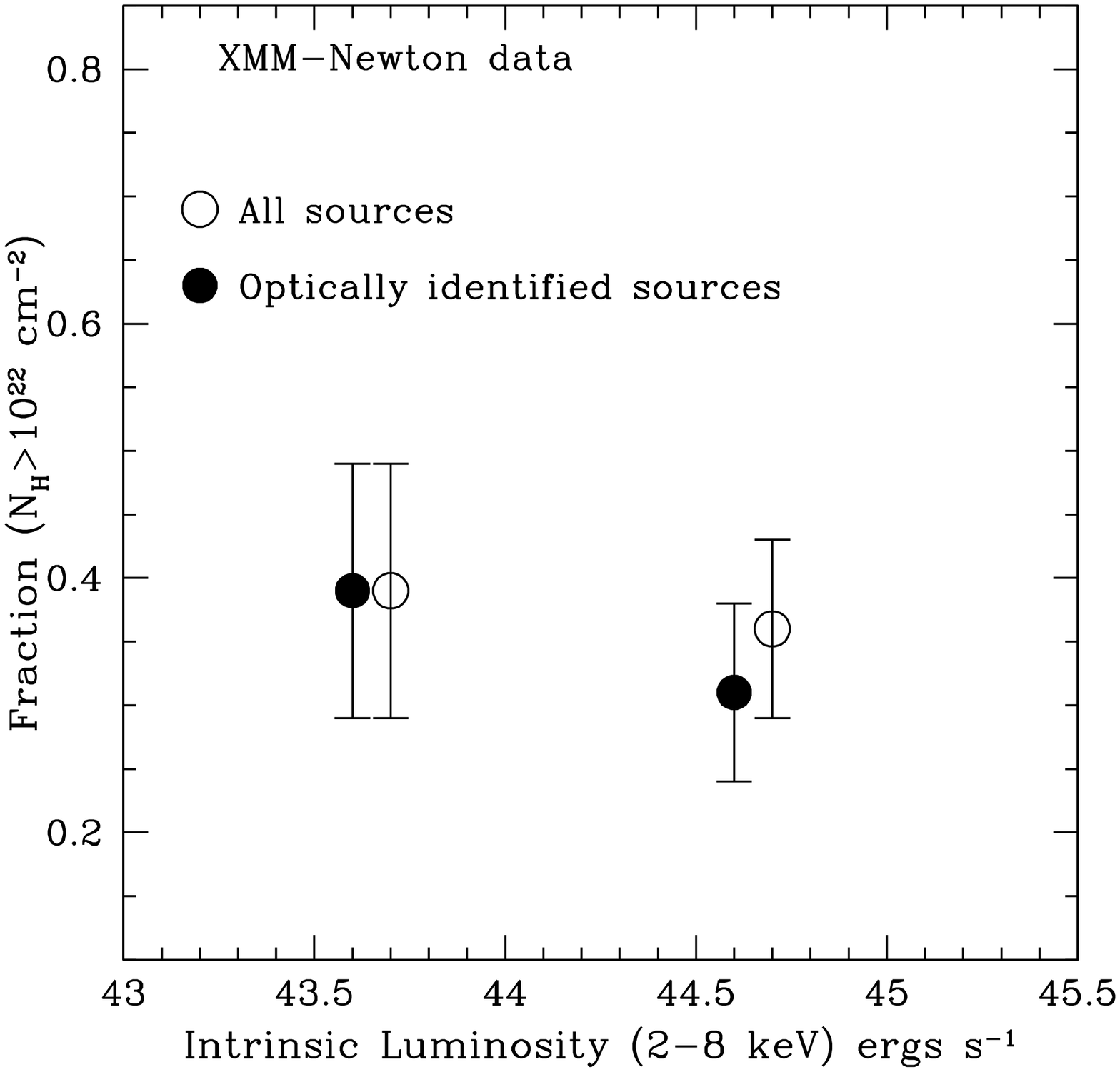}
   \includegraphics[height=6.8cm]{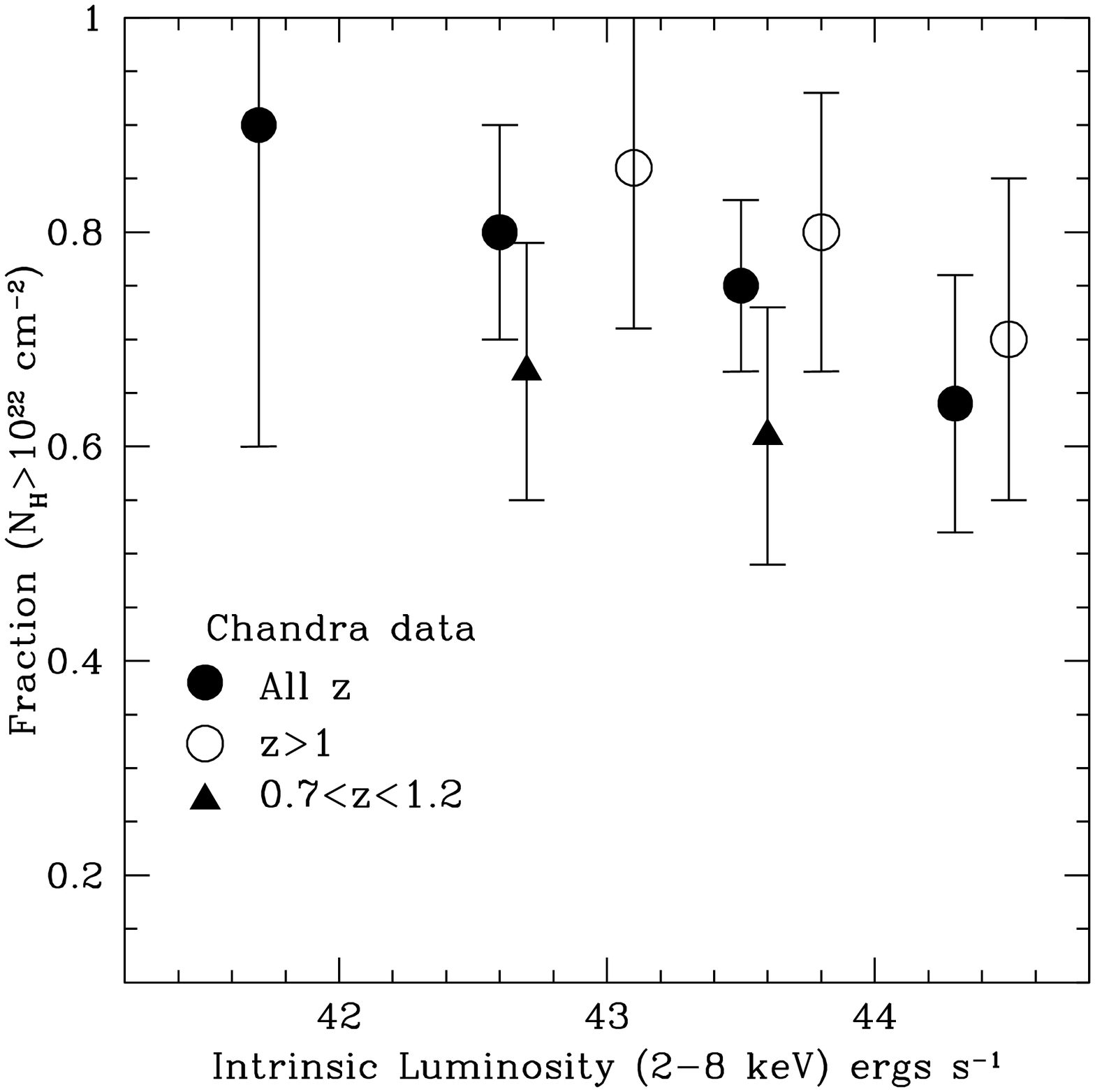}
   \includegraphics[height=6.8cm]{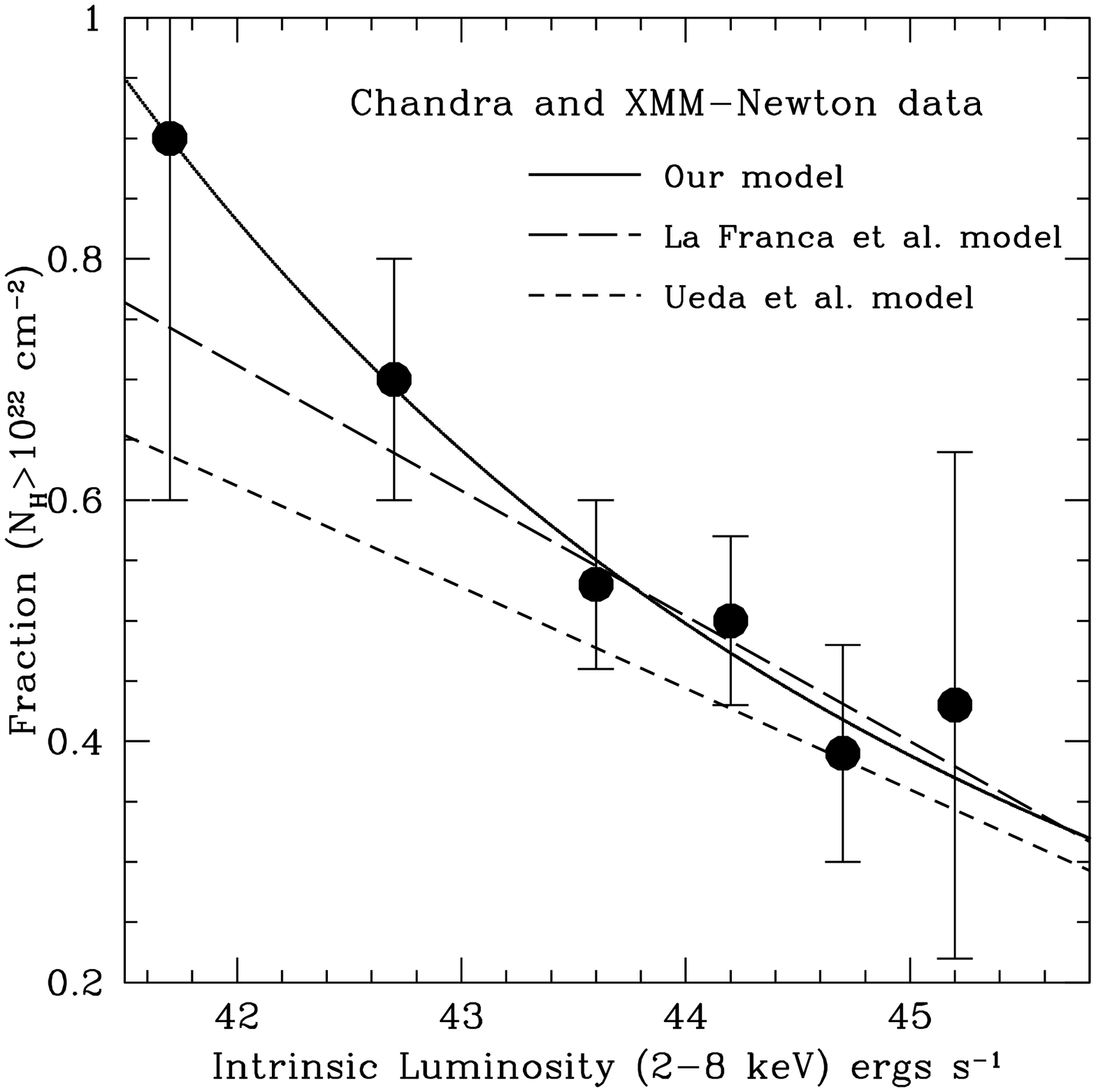}
   \caption{The fraction of obscured ($N_H>10^{22}$ cm$^{-2}$) 
        sources presented in the {\xmm} data (upper panel) and the 
	\chandra data (middle panel) as a function of the 2-8 keV, 
        intrinsic luminosity. In the lower panel we plot the same 
	fraction for the combined {\xmm} and the {\chandra} samples. 
	In this panel we also plot our best fit model as well as of 
	Ueda et al. (2003) and La Franca et al. (2005).}
   \label{vmaxl}
   \end{figure}

In Fig. \ref{vmaxl} we plot the estimated fraction of 
absorbed sources ($N_H>10^{22}$ cm$^-2$) in a certain 
luminosity bin as a function of the  median luminosity 
of this bin for the {\xmm} (upper panel). 
When we use only the 171 optically identified sources there 
is a marginal reduction in the fraction of obscuration at higher 
luminosities ($>10^{44}$ \lunits). This weakens even further
when we include the 26 optically faint sources 
(assuming z=1.5). This is reasonable since these sources are 
most probably associated with obscured sources. 

%In the case of the {\chandra} data the adopted luminosity bins 
%are $L_X<10^{42}$ \lunits, $10^{42}<L_X<10^{43}$ \lunits, 
%10^{43}<L_X<10^{44}$ \lunits, $10^{44}<L_X<10^{45}$ \lunits. 
In the case of the CDF-S data (middle panel in Fig \ref{vmaxl}) 
there is again a marginal but constant decrease of the 
fraction F with increasing luminosity.
In a flux limited sample it is reasonable to expect a strong 
correlation of luminosity on redshift (see Fig. \ref{lz}). 
We attempt to  break this degeneracy by 
exploring in Fig. \ref{vmaxl} (middle panel, filled triangles) 
the fraction of obscured \chandra sources as a function of 
luminosity in a thin redshift slice ($0.7<z<1.2$). This slice 
is chosen so as to maximize the number of objects.   
%It appears that the fraction falls with 
%luminosity suggesting that this effect 
%is not induced by redshift. 
Finally, in the same panel we plot  the fraction versus  
luminosity for the high redshift ($z>1$) \chandra sources. 
The purpose of this is to test whether the fraction of 
absorbed sources increases at higher redshift.  Indeed the K-correction
shifts the curvature of the X-ray spectrum caused by the absorption 
at low energies, i.e. reduces  the absorption measured 
at the observer's frame. This should result in an 
increased number of absorbed sources at higher redshift. 
A marginal decrease of F with increasing luminosity exists 
in all three cases.

%   \begin{figure}
%   \centering
%   \includegraphics[width=8.5cm]{vmax_nh_z.ps}
%   \includegraphics[width=8.5cm]{vmax_nh_z.chandra.ps}
%   \caption{The fraction of obscured ($N_H>10^{22}$) sources
%    presented in the {\xmm} data (upper panel) and the 
%   \chandra data (lower panel) as a function of the redshift.}
%   \label{vmaxz}
%   \end{figure}

However the small L-z plane coverage of the individual 
samples does not allow us to provide strong contrains
on the F-L dependence.
Therefore in  order to explore the widest luminosity and redshift 
range we combine the {\xmm} and the {\chandra} datasets. 
Thus we obtain a catalogue comprising 359 sources in the 
2-8 keV flux range of $6\times10^{-16}$ to $4\times10^{-13}$ \funits. 
In Fig. \ref{vmaxl} (lower panel) we plot the fraction of 
the obscured sources for 
the combined ({\xmm} and {\chandra}) sample obtained using the 
$1/V_{m}$ method. 
The combined data reveal a clear decrease in 
the fraction of absorbed sources at higher luminosities. 
The solid line shows the best fit model to the data. 
The fraction of absorbed sources 
is related to the luminosity according to 
$F(N_H>10^{22} {\rm cm^{-2}})=7.075\times10^{17}(LogL_X)^{-11.045}$.
In the same panel in Fig. \ref{vmaxl} we also plot the 
best fit models presented in Ueda et al. (2003) 
and La Franca et al. (2005).

% In Fig. \ref{vmaxz} we plot the fraction
% of absorbed sources as a function of redshift
% for the \xmm (upper panel) and \chandra (lower panel). 
% In this case the estimate of the maximum volume 
% for each source, and in 
% each redshift bin, is restricted by the lower and 
% higher redshift value of that bin.
% In the \xmm data we observe an increase 
% of the fraction with redshift. 
% The same trend is observed for the \chandra data 
% apart from the lowest redshift bin ($z<0.25$).  
 In Fig. \ref{red_model}, we plot the fraction F 
 against redshift for the combined \xmm and 
 \chandra data (the optically unidentified 
 {\xmm} sources are excluded).
  {It is important here to note that in this plot we are plotting the 
  'observed' fraction as a function of redshift, i.e. 
  we have not applied any $1/V_{m}$ correction to the data. 
  This allows a direct  comparison  with the models which 
  predict the expected fraction of sources with redshift for a 
  given survey.

  The dashed-dotted line model shows the expected fraction using the 
 Ueda et al. (2003) luminosity function in the case 
 where there is no dependence of the obscured fraction 
 on luminosity. A value of R=1 for the ratio of 
 obscured to unobscured AGN has been used. 
 Note that the ratio R is related to the fraction F 
 according to the relation F=R/(1+R). 
 In this case the model predictions increase with redshift because 
 of the  K-correction, i.e. more obscured sources are 
 detected as the column decreases at higher redshifts 
 according to $(1+z)^{2.65}$.
 The solid line model describes the expected fraction  
 of absorbed sources using the luminosity function of 
 Ueda et al. (2003), combined with our best fit $F-L_X$ 
 relation derived previously. The effect of the steep 
 $F-L_X$ relation is to roughly cancel out the K-correction 
 effect. The data show a significant increase in the 
 fraction of obscured objects, F, at higher redshifts (z$>$2). 
 This behavior, if not real, may be caused by the  
 photometric redshift estimations or alternatively by small 
 fluctuations in the lowest energy bins which 
 translate to a significant absorption at high z.
 In order to test the first possibility we repeat the 
 calculations considering the photometric and the spectroscopic data  
 separately. The resulting plots are very similar and  still there 
 is an  increase of the fraction of obscured sources at high z. Therefore
 we conclude that the photometric redshifts cannot introduce this trend.  
 We examine the significance of the second effect by using 
 spectral simulations (see Appendix A for details). 
 In Fig. \ref{red_model} the long dashed line shows the input 
 distribution of the obscured sources used in the simulations and 
 the sort dashed line the resulting one after fitting the simulated 
 spectra in {\sl XSPEC}. Clearly appears an increase in the fraction F
 which is solely introduced by some fluctuations in the 
 lowest energy bins of the spectral files. 
 This suggests that the observed fraction of obscured sources at high $z$ 
 is artificially enhanced. 
 Consequently there is no significant evidence for an increase in the 
 fraction F with redshift.

Next we investigate the evolution
of the AGN space density in different 
luminosity bins as a function of redshift. 
For this analysis we consider only the CDF-S data
in order to use a sample with complete redshift information. 
The estimation of the space density is 
based on the 1/V$_{m}$ method. The space 
densities as a function of redshift are 
calculated in four luminosity bins in the ranges 
$\rm Log(L_X)$ 42-43, 43-44, 44-44.5 and 44.5-45.5
\lunits. The results are plotted in 
Fig. \ref{density} as a function of the median 
redshift of each redshift bin. 
The 1$\sigma$ errors are also plotted. 
Fig. \ref{density} clearly shows a shift of the number 
density peak with luminosity in the sense that
more luminous AGN peak at an earlier era, while the less
luminous ones arise later. There is also  evidence for a decline 
in the density of the lower  luminosity 
QSOs (LogL$_X$$<$44 \lunits) and especially those at  
42$<$LogL$_X$$<$43. 
This trend is known as cosmic down-sizing and it has 
previously being reported by Ueda et al. (2003),  Fiore et al. (2003) and 
Barger et al (2005). Hasinger et al. (2005), also found similar 
results, analyzing the space density of type-I QSOs.
     
 \begin{figure}
 \centering
 \includegraphics[width=8.5cm]{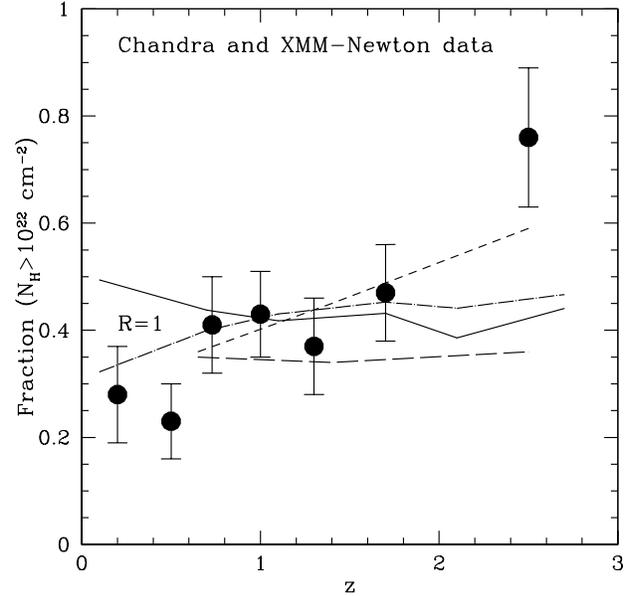}
 \caption{The {\it observed} fraction of obscured AGN as a 
 function of redshift. The solid line gives the expected
 fraction assuming the luminosity function of Ueda
 et al. (2003) combined with our best fit model relation between 
 the obscured fraction of AGN and luminosity. The dotted-dashed 
 line gives the expected fraction assuming no dependence of F on 
 luminosity for a ratio of obscured to unobscured AGN R=1. 
 The long dashed line shows the input distribution of the 
 fraction of obscured sources used in our simulations 
 (see Appendix A) and the sort dashed line the resulting one after fitting the 
 simulated spectra.}
 \label{red_model}
 \end{figure}

%In particular for the {\xmm} data (see the upper panel 
%in Figure \ref{vmaxz}) we consider the fraction of absorption
%in only two redshift bins, $z<0.7$ and $z\geq0.7$ with 
%comparable number of sources.
% We also examine the case where the unidentified
%sources are included in our sample.
%In the case of the \chandra data, we consider four redshift 
%intervals, $z<0.7$, $z\geq0.7<1.1$, $z\geq1.1<2.0$, 
%$z\geq2.0$ (see the lower panel in Figure \ref{vmaxz}). 

   \begin{figure}
   \centering
   \includegraphics[width=8.5cm]{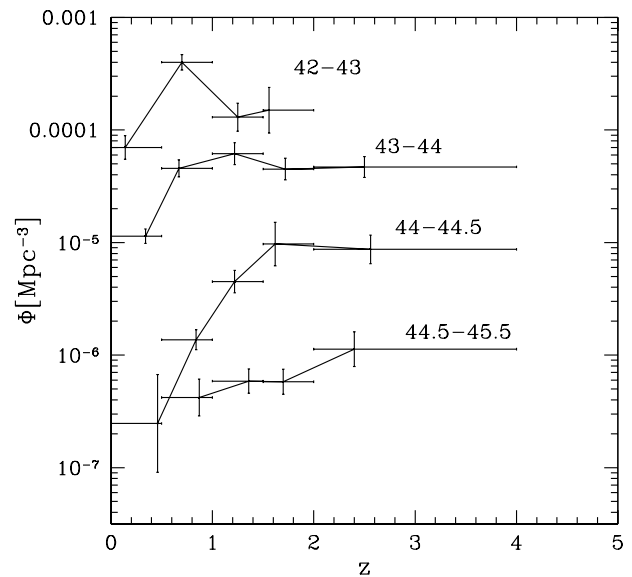}
     \caption{The space density of the CDF-S AGN as a function of redshift in 
     four luminosity ranges, Log$L_X$=42-43, Log$L_X$=43-44, 
     Log$L_X$=44-44.5 and Log$L_X$=44.5-45.5. The errors correspond 
     to the 1$\sigma$ confidence level.}
    \label{density}
    \end{figure}

\begin{table}
\centering 
\caption{The median values for the  2-8 keV observed flux for 
each luminosity bin in the combined sample.}
\label{medianf2}
\begin{tabular}{cccc}
\hline
${LogL_X}^1$ &     ${f_X}^2$      \\
\hline			      
\hline			      
42.7   & $1.7\times10^{-15}$   \\
43.6   & $5.3\times10^{-15}$   \\
44.5   & $3.1\times10^{-14}$   \\
45.2   & $5.6\times10^{-14}$   \\
\hline
\end{tabular} 	    
\begin{list}{}{}
\item$^1$ Logarithm of the median value of 2-8 keV in units of \lunits.
\item$^2$ Median value of the 2-8 keV observed flux in units of \funits.
\end{list}
\end{table}

\section{DISCUSSION}

 In this paper we use the largest {\xmm} 
 sample with X-ray spectroscopic information available
 in order to investigate the  behavior of 
 intrinsic absorption in AGN  
 as a function of redshift and luminosity. We also 
 take  advantage of the complete optical coverage 
 (photometric or spectroscopic) of the CDF-S observations 
 in order to extend our results at fainter fluxes 
 (down to $\sim6\times 10^{-16}$ \funits). 

In Fig. \ref{vmaxl} we  plot the fraction of obscured objects 
as a function of luminosity separately for the {\xmm} 
(upper panel), the {\chandra} data (middle panel) and their 
combination (lower panel).
Despite the weak indications for a decline of this fraction 
in the individual samples, the only way to unambiguously verify 
this correlation is to increase the coverage of the L-z plane by combining 
the two samples.
% In the first case there is a marginal evidence for a decline of 
% this fraction as we move toward brighter luminosities. 
% This trend is more obvious in the case of the \chandra data. 
% We find that this trend persists when we plot only a thin 
% redshift slice ($0.7<z<1.2$) and hence it appears that this 
% effect is not  induced by a redshift-luminosity correlation. 
% When we combine the \xmm and \chandra data the fraction of 
% obscured objects falls dramatically with increasing luminosity. 
It is likely that the rapid decline appearing in the combined sample 
could not be easily observed in the {\xmm} and the {\chandra} individual 
datasets due to the limited available volume of each independent, flux 
limited survey. For example {\chandra} does not cover a large enough 
volume to sample a large number of luminous sources. However, 
we caution that the combination of these different subsamples may 
introduce some bias in favor of an $F-L_X$ correlation. 
Indeed, as we move toward higher luminosities we sample more 
{\xmm} sources and less {\chandra} sources in each luminosity bin 
according to Fig. \ref{lz}. But the {\xmm} sources are found in much 
brighter fluxes in comparison with the {\chandra} ones. Thus, as we 
move progressively to higher luminosities we sample higher fluxes and 
hence less obscured sources according to the well known flux-absorption 
correlation. This effect is summarized in Table \ref{medianf2} were we 
list the median 2-8 keV flux for each luminosity bin presented 
in Fig. \ref{vmaxl}. 

The decrease  of the fraction of obscured sources as a function of 
luminosity is consistent with the results of La Franca et al. 
(2005). The physical interpretation of this model could be that 
the radiation pressure flattens the torus in luminous objects 
(K\"{o}nigl \& Kartje 1994) or increases the degree of photoionization 
of the gas around them. Another possible scenario is that of the 
'receding torus' which has been proposed by Lawrence (1991) and has 
been recently updated by Simpson (2005), where because of the effects of 
dust sublimation the inner radius of the torus increases with luminosity.

One way to explore whether the relation between the absorbed 
fraction and luminosity ($F-L_X$ relation) is real (or is induced 
up to some degree by the strong flux-absorption correlation) 
is to model the number of absorbed sources as a function of flux. 
In Fig. \ref{xrbmodel} we plot the fraction of obscured AGN, F, 
as a function of the flux. 
We plot separately the {\xmm} points, the CDF-S points 
as well those from the {\chandra} survey in the 
Extended Groth Strip (Georgakakis et al. 2006b) and 
compare with various model predictions. 
In the above models we use the Ueda et al. (2003) 
luminosity function. The solid line gives the predictions
of our model and uses the $F-L_X$ relation derived here. 
The long dashed line corresponds to a model with a ratio of 
obscured to unobscured AGN, R=4 (or F=0.8) with no 
dependence on luminosity. This is the ratio derived 
in the local Universe by Risaliti et al. (1999) 
and Maiolino \& Rieke (1995). The short dashed line 
corresponds to the R=1 case with no dependence on luminosity.  
The R=1 model nicely represents the {\xmm} data while the R=4 
model follows better the faint CDF-S data. However, only a 
model which includes a decrease in the $F-L_X$  relation can 
explain the abrupt increase of the fraction of absorbed sources 
with decreasing flux.

Previous estimates are in agreement with  
our results. Piconceli et al. (2003), analyzing hard 
X-ray {\xmm} data, claimed that the observed fraction of 
obscured sources at bright fluxes ($>10^{-13}$ \funits) 
is about 30 per cent, much lower than that predicted by 
the (R=4) XRB model.  La Franca et al. (2005), combining 
data from different X-ray samples, studied the behavior 
of obscuration in a much wider flux range. Their findings
for the fraction of obscured sources are very similar to ours
in both faint and bright fluxes. 
Their proposed $F-L_X$ relation is in agreement with our 
model (see Fig. \ref{vmaxl}). Note that here we have used 
a power-law best fit model instead of a linear one.  
 
Fig. \ref{red_model} shows the dependence of $F$  
on redshift. 
The fraction shows an apparent increase with redshift, with
a more abrupt increase at high redshifts, $z>2$. 
Ueda et al. (2003) find no dependence of the fraction $F$ 
on redshift. Ballantyne et al. (2006) propose that 
the obscuration may be related to star-formation within 
the host galaxy and thus there may be some increase of the 
obscured AGN fraction  with redshift.
They test their models by comparing with the 
observed type-I AGN in the CDF-N (Barger et al. 2003).
These authors find an obscured  fraction F evolving as $(1+z)^{0.3}$,
together with a dependence of F on luminosity. 
La Franca et al. (2005) find a dependence on redshift very 
similar with ours (see their Fig. 6 right panel): 
the fraction $F$ increases from $\sim0.2$ at low redshift 
to $F\sim0.6$ at $z>2$. In our case however, our simulations 
show that the lowest energy bins fluctuations can introduce
a significant artificial correlation of the fraction F with 
redshift. This effect becomes particularly important at 
high redshifts due to the K-correction effect. 
This suggests that the observed fraction of obscured sources at high $z$ 
is erroneously  enhanced. 
Consequently there is no significant evidence for an increase in the 
fraction F with redshift.

   \begin{figure}
   \centering
   \includegraphics[width=8.5cm]{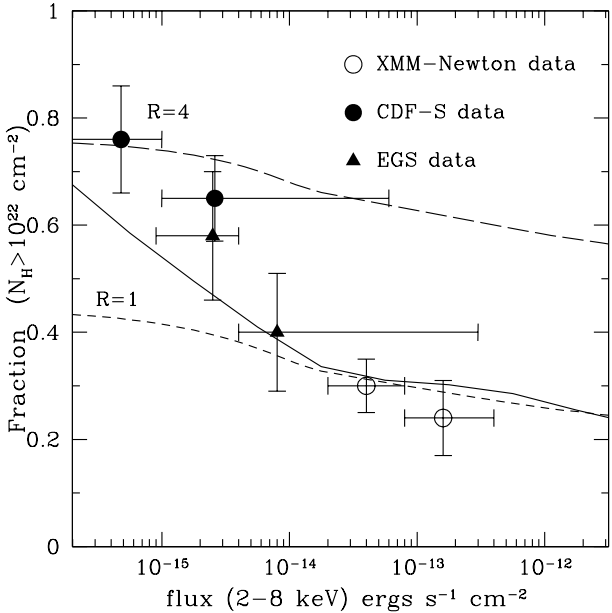}
     \caption{The fraction of the sources with $N_H>10^{22}$ cm$^{-2}$
	for the {\xmm} sample (open circles) and the 
	{\chandra} sample (filled circles). The triangles
        represent the EGS \chandra survey of Georgakakis 
 	et al. (2006b). The solid line gives the  predicted 
	fraction using the $F-L_X$ relation derived here. 
	The long and short dashed lines give the models with 
 	R=4 and R=1 ratios of obscured to unobscured AGN 
 	respectively with no dependence on luminosity.}
    \label{xrbmodel}
    \end{figure}

\section{SUMMARY}
 We have combined bright {\xmm} (from SDSS fields and the XMM1dF survey)
 and faint \chandra (from CDF-S)  data to form the largest 
 sample (359 sources in the 2-8 keV band)
 with X-ray spectra. Our goal is to 
 investigate the intrinsic AGN obscuring column density    
 as a function of luminosity and redshift. 
 This bears important implications on both AGN 
 unification models as well as the X-ray background 
 population synthesis models. 
 The CDF-S  has complete redshift coverage ensuring 
 that there is no bias because of optically 
 unidentified sources at faint fluxes. At bright fluxes the 
 level of redshift incompleteness is less than 15 per cent. 
 We use the $1/V_m$ method to estimate the fraction, F,
 of obscured to unobscured AGN. This properly takes into
 account the bias introduced by the fact that 
 obscured sources are fainter in flux and thus 
 are preferentially detected in smaller numbers 
 and at preferentially lower redshifts. 
 Our findings can be summarized as follows:

 The fraction of obscured AGN, F 
 decreases with increasing luminosity. This confirms 
 previous results by Ueda et al. (2003) and La Franca 
 et al. (2005). The dependence of the fraction F on 
 luminosity naturally reproduces the observation 
 that the number of obscured sources increases 
 drastically with decreasing flux.
 
 There is tentative evidence for a increase 
 of the fraction F with increasing redshift. 
 However, this is  mainly based on the high redshift 
 bins ($z>2$) and thus should be viewed 
 with caution. Our simulations show that these bins
 are affected by systematic overestimates of the column density, 
caused by small fluctuations in the lowest energy spectral bins.

\begin{acknowledgements}
  This work has been supported by the program  
 Promotion of Excellence in Research, 
 grant `X-ray Astrophysics with ESA's mission XMM' 
 funded jointly by the European Union 
 and the Greek Ministry of Development.
  This work is based on data obtained
 from the {\xmm} and CXC public data archives.
 We acknowledge the use of data from the 
 Sloan Digital Sky Survey. 
 \end{acknowledgements}

\appendix
\section{Validity of the $N_H$ estimations: Spectral simulations  }
We describe here the strategy we follow in order to check the validity of 
the derived $N_H$ estimations through the spectral analysis. 
Given that a small fluctuation in the lowest energy bin in our \chandra spectral 
files might correspond to an apparently high absorption at high $z$ we have tried some 
simulations to quantify this effect. 
First we construct 920 fake spectral files with exactly the same redshift and flux 
distribution suggested from the SDF-S data. Each spectral file presents an arbitrary
amount of absorption in the range of 10$^{20}$-10$^{23}$ cm$^{-2}$ and a photon index 
of 1.8. These fake files are fitted in XSPEC in order to obtain a value for the 
observed column density. The obtained values are converted to the intrinsic 
$N_H$ value based on the formula ${N_H}_{intr}=({N_H}_{obs}-{N_H}_{gal})\cdot(1+z)^{2.65}$.
The results are plotted in Fig. \ref{sim}. The solid line histogram describes the initial $N_H$ 
distribution of all the sources and the short dashed line the corresponding one after 
fitting the fake data in XSPEC. Also the dotted line histogram shows the initial distribution 
of the low redshift ($z<1$) sources and the sort dashed line the resulting $N_H$ distribution
for the same population. Clearly the spectral fitting analysis produce 
a significant increase in the number of obscured ($N_H>10^{22}$ cm$^{-2}$) sources in comparison with the 
initial distribution. The same plot suggests that this effect is almost negligible in the low $z$ 
population. Our conclusion is that high $z$ sources suffers from a systematic increase in the 
measured column density. This overestimate is about 50 per cent at $z\sim$2.5, 20 per cent at 
$z\sim$1.5, and become almost negligible at $z<1$.  

   \begin{figure}
   \centering
   \includegraphics[width=8.5cm]{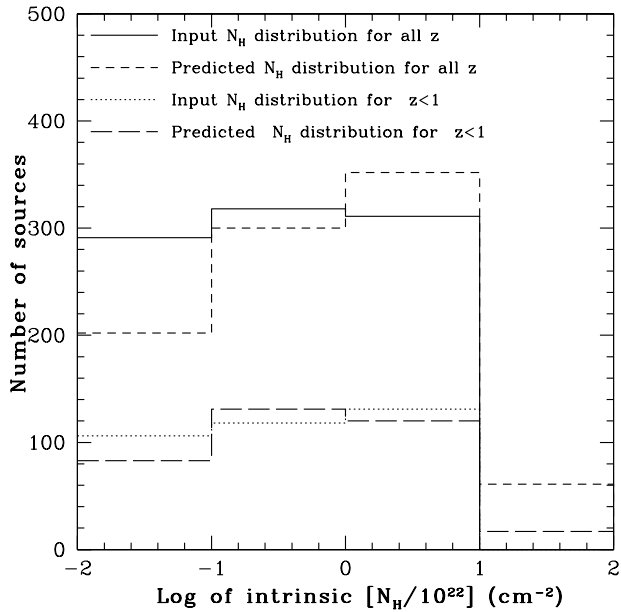}
     \caption{This plot summarize the results of the performed 
	simulations concerning the validity of the spectral analysis of 
	the CDF-S data. The solid line histogram describes the initial $N_H$ 
	distribution of all sources and the short dashed line the 
	corresponding one after fitting the faked data in XSPEC. 
	Similarly the dotted and the long dashed histograms show the same 
	results, using only the low $z$ ($z<1$) data.}
    \label{sim}
    \end{figure}

\end{document}